\documentclass{aa}  
\usepackage{graphicx}
\usepackage{soul}
\usepackage{color}
\usepackage[colorlinks=true,     linkcolor=blue, citecolor=blue, filecolor=blue, urlcolor=blue]{hyperref}
\usepackage[]{natbib} 
\usepackage{txfonts}

\newcommand{\mi}{$\mu$m}

\newcommand{\mm}{M\,33}
\newcommand{\Ha}{H$\alpha$}

\begin{document}

   \title{Star formation drivers across the M33 disk}

   \author{Edvige Corbelli
          \inst{1}
          \and
          Bruce Elmegreen\inst{2}
          \and
          Sara Ellison\inst{3}
          \and
          Simone Bianchi\inst{1}
          }

   \institute{INAF-Osservatorio Astrofisico di Arcetri, Largo E. Fermi, 5,
             50125 Firenze, Italy\\
             \email{edvige.corbelli@inaf.it} 
         \and
             Katonah, NY, USA\\
             \email{belmegreen@gmail.com}
             \and
             Department of Physics $\&$ Astronomy, University of Victoria, Finnerty Road, Victoria, British Columbia, V8P 1A1, Canada\\
             \email{sarae@uvic.ca}
              }

   \date{Received ...; Accepted ....}
   
 
  \abstract
   { }
   {We investigate  the star formation process across M33, which is characterized by a low molecular content and can be sampled with high spatial resolution out to regions where star formation activity drops.
 }
   {We used a multiwavelength dataset  and disk dynamics to
   extract the local physical parameters across the M33 disk, such as the atomic, molecular, stellar and dust mass surface densities, dark matter densities, and hydrostatic pressure. We computed numerically  equilibrium values of gas densities and scale heights across the disk,  testing several analytic approximations that are often used to estimate these variables. Orthogonal regressions and hierarchical Bayesian models, as well as random forest (RF) analyses, were used to establish the fundamental relations at physical scales from  160~pc to 1~kpc. 
   }
   {The gas hydrostatic pressure, P$_{hy}$, which balances the local weight, is the main driver of the star formation rate surface density, $\Sigma_{SFR}$, throughout the whole star-forming disk of M33. High-pressure regions enhance the atomic-to-molecular gas conversion, with the molecular hydrogen mass surface density, $\Sigma_{H_2}$, being tightly correlated to P$_{hy}$ and a uniform scaling law throughout the M33 disk.  The  P$_{hy}$--$\Sigma_{SFR}$ relation differs, showing a change in slope  from the inner to the outer disk. Our use of an accurate analytic expression and database to compute P$_{hy}$ for a multicomponent disk minimizes observational scatter. This  points to scaling laws  that do not  depend on the physical scale  and brings out an intrinsic scatter linked to variations in the efficiency and relative age of the  molecular gas-to-stars conversion. In the inner disk, where spiral arms are present and the stellar surface density dominates gravity, P$_{hy}$ and $\Sigma_{SFR}$ establish an almost linear correlation with a smaller dispersion than in the $\Sigma_{H_2}$--$\Sigma_{SFR}$ relation. In the atomic gas-dominated outer disk, $\Sigma_{SFR}$ has a steeper dependence on P$_{hy}$,  which we propose could be the result of an increasing fraction of diffuse molecular gas  that does not form stars.
   }
   {}
   

   \maketitle


\section{Introduction}

The study of star-forming regions in our Milky Way has probed  key physical processes that drive and regulate the conversion of molecular gas into stars \citep{2007ARA&A..45..565M,2012ARA&A..50..531K}. Understanding star formation on a larger scale is also essential for framing galaxy evolution. Star formation is, in fact, regulated by a combination of processes \citep{2014PhR...539...49K}, some very local, such as the interstellar medium (ISM) turbulence, cooling, and gravitational collapse, and others related to the disk dynamics and morphology, such as density waves, bars, and large-scale instabilities.  Whether star formation is more controlled by the local conditions or by disk dynamics is still an open question. One local model of star formation  that is emerging from numerical simulations is the pressure-regulated, feedback-modulated (PRFM) theory \citep[][hereafter OK22]{2022ApJ...936..137O}, in which stellar feedback  sustains the thermal and dynamical equilibrium of the ISM. 

Scaling laws that relate physical quantities are a first step to improve our understanding of the key ingredients for star formation. Tracing these laws, which relate the global and local properties to the star formation rate (SFR), is now becoming possible thanks to multiwavelength surveys of resolved galaxies. The EDGE-CALIFA survey for example  \citep{2016A&A...594A..36S,2017ApJ...846..159B,2024ApJS..271...35W} and the ALMA MaNGA QUEnching and STar formation (ALMaQUEST) survey aim to understand the kpc-scale processes that regulate star formation in the nearby universe  \citep{2020ApJ...903..145L,2024MNRAS.52710201E}. The Physics at High Angular resolution in Nearby GalaxieS (PHANGS)  survey provides the opportunity to  explore scaling relations at 150 pc resolution, although the use of data at long wavelength often requires a resolution of order 1.5 kpc \citep{2021ApJS..257...43L,2022A&A...659A.191E,2022AJ....164...43S}. The galaxy  environment is also able to affect the ability of the gas to form stars through gas removal mechanisms or compression as galaxy-galaxy interact  \citep{2010A&A...518L..48D,2018ApJ...855....7H,2018A&A...614A..56B,2021ApJS..257...21B, 2021PASA...38...35C,2023A&A...671A...3J,2023ApJ...956...37B}.

The most common scaling relation is between 
the SFR surface density, $\Sigma_{SFR}$, and gas, $\Sigma_{gas}$, or molecular gas surface density, $\Sigma_{H_2}$,  known as the resolved Schmidt-Kennicutt relation (KS) \citep[][and references therein]{2012ARA&A..50..531K}, with a similar correlation between the SFR and the dense gas \citep{2004ApJ...606..271G,2025A&A...693L..13N}. These relations  indicate that star formation is driven by the abundance of cold gas. When the total gas column density is measured, the correlation between $\Sigma_{SFR }$ and $\Sigma_{gas}$   is superlinear.  For molecular gas it is close to linear. Both relations are shallower if the  gas column density is divided by a dynamical timescale, orbital, or free-fall \citep{2002ApJ...569..157W,2008AJ....136.2846B,2011AJ....142...37S,2023ApJ...945L..19S,2024MNRAS.52710201E}. 

Integral field spectroscopy (IFS) using large samples of star-forming galaxies has also related $\Sigma_{SFR}$  to the stellar surface density $\Sigma_*$  at kpc scales. 
This scaling relation, known as the resolved star-forming main sequence \citep[SFMS][and references therein]{2020ARA&A..58...99S}, may be understood in terms of the stellar contribution to local gravity. The presence of a SFMS therefore leads us to expect an underlying relation between ISM pressure and $\Sigma_{SFR}$. Indeed, since the pioneering work of  \citet{1993ApJ...411..170E}, \citet{2002ApJ...569..157W} and \citet{2004ApJ...612L..29B} several recent studies have found a tight relation  between  $\Sigma_{SFR}$ and various formulations of the local pressure, such as the thermal pressure, P$_{th}$ \citep{2017ApJ...835..201H}, hydrostatic pressure, P$_{hy}$, or  dynamical equilibrium pressure, P$_{DE}$ \citep{2021MNRAS.503.3643B,2024MNRAS.52710201E,2020ApJ...892..148S,2024ApJ...966..233E,2024A&A...691A.163E}.  The interplay between $\Sigma_{SFR}$, $\Sigma_*$, and $\Sigma_{H_2}$ (and how it depends on the galactic environment) has also been examined for PHANGS galaxies at a spatial resolution of 150~pc \citep{2022A&A...663A..61P}. Azimuthal fluctuations have also been examined to avoid mixing radial dependencies with scaling laws of physical quantities,  and  they suggest that both feedback and galactic dynamic play a role \citep{2024ApJ...966..233E}. 

Using the extended ALMaQUEST sample, \citet{2024MNRAS.52710201E} found that the relation between $\Sigma_{SFR}$ and P$_{DE}$ for starburst galaxies that are at least a factor of 2 above the SFMS is flatter than predicted by the PRFM theory. Nonlocal drivers of star formation, such as inflows of gas, may explain the observed  deviation \citep{2018MNRAS.477.2716K}.
However, most studies with large samples, such as ALMaQUEST, are limited in spatial resolution  and no information on the atomic gas surface density and dispersion is available.

Here, we examine our nearest undisturbed star-forming galaxy M33, testing star formation drivers from the center to the edge of the star-forming disk.  M33 gives us the opportunity to examine spatial resolution effects as well as assumptions sometimes used in deriving physical quantities for less resolved galaxies or with incomplete multiwavelength maps. M33 shows no evidence of ongoing or past interactions \citep{2017MNRAS.464.3825P,2019ApJ...872...24V,2024A&A...685A..38C}, no bulge, and a very small, weak bar \citep{2007ApJ...669..315C}. It is dominated by atomic gas with HI three times more extended than  the star-forming disk, in which a small fraction of the gas is molecular \citep[about 10$\%$;][]{2014A&A...567A.118D}. It has a gas mass only half the stellar mass  and it is viewed more face-on compared to M31, the only other spiral galaxy in the Local Group.  The ISM at sub-solar metallicity stellar populations and disk dynamics of M33 have been widely traced with high resolution and sensitivity \citep[e.g.,][]{2009A&A...493..453V,2010A&A...512A..63M,2011A&A...533A..91G,2014A&A...572A..23C,2014A&A...567A.118D,2024PASJ...76.1098K}. 
Dedicated numerical simulations have also shown how stellar feedback and gravitational instability play a crucial role in setting the ISM morphology \citep{2018MNRAS.478.3793D} although the relative importance of these processes might change across the disk. 
There is also a stellar component associated with the outer disk, beyond the region of current star formation, namely, 100-300~Myrs ago \citep{2011A&A...533A..91G}. 
The study of the star formation in this galaxy is now reaching unprecedented detail thanks to  high resolution observations from UV to radio wavelengths \citep[e.g.][]{2020ApJ...896...36T,2024MNRAS.52710668P}. These are paving the way to a comprehensive view of  the star formation process in a galaxy different than the Milky Way. All this makes M33 an ideal laboratory for investigating the role of feedback, disk dynamics, and environment in triggering the formation of stars. 

The paper plan is the following. In Section 2, we describe the data and methods used to extract physical quantities. In Section 3 and in Appendix~\ref{appsh}, we describe how we tested  simple analytic recipes to derive the local midplane  pressure and the stellar and gaseous disk scale heights. Radial variations, used to define the inner and outer disk, are shown in Appendix~\ref{apprad}. Star formation drivers across the disk are analyzed in Section 4, with Appendix~\ref{appfuv} referring to possible variations of the working assumptions. Section 5 summarizes the results presented in this paper and our conclusions.

\section{Data}

In analyzing the data of M33, we assumed a distance of 840~kpc as resulting from cepheid variables  \citep{1991ApJ...372..455F,2001ApJ...553...47F,2013ApJ...773...69G}, which implies a linear scale of 4.1~pc per arcsecond.  The inclination and position angle used for the M33 disk are  54$^\circ$ and 22$^\circ$, respectively,  in close agreement with the values inferred from disk dynamics and from isophotal photometry in B-band \citep{1991rc3..book.....D,2014A&A...572A..23C}. All mass surface densities are face-on values and all gas masses include also corrections for heavy elements. The results and considerations expressed in Appendix~\ref{apprad} support the definition of the inner star-forming disk for R$\le 4$~kpc and the outer star-forming disk for $4<$R$\le7$~kpc.

\subsection{Multiwavelength maps }

All  M33 emission maps  used in this paper  have been aligned with the galaxy optical center and background subtracted if they refer to the continuum emission. We use the maps at their original spatial resolution, which varies between 2~arcsec  for the H$\alpha$ map to 12~arcsec for the $^{12}$CO J=2-1 map. 

To investigate the ultraviolet (UV) continuum emission of \mm, we used 
{\it Galaxy Evolution Explorer (GALEX)} data, in particular, those distributed by \citet{2007ApJS..173..185G} in the far- and near-ultraviolet bands (FUV and NUV). 
To trace the ionized gas, we adopted the narrow-line \Ha\ image of \mm\ obtained by \citet{1998PhDT........16G}
and described in detail in \citet{2000ApJ...541..597H}. 
The H$\alpha$ and FUV surface brightnesses in M33 decrease radially with a scale length of about 2~kpc out to about 6.5~kpc
\citep{2009A&A...493..453V}. Beyond this radius, they experience a sharper radial decline. 

The distribution of  the neutral atomic gas and of its dispersion along the line of sight, $\sigma_{HI}$ traced by the 21cm line emission in M33 has been mapped by combining  
Very Large Array (VLA) and  Green Bank Telescope (GBT) data and the results have been described by \citet{2014A&A...572A..23C}. We used the mom-0 and mom-2 maps at 10~arcsec spatial resolution resolution.  The sensitivity of this map  is enough to detect the 21-cm line surface brightness throughout the whole star-forming disk. This map has been recovered from spectral cube with 1.25~km~s$^{-1}$ channel width and has a higher spatial resolution than  the more recent  VLA map \citep{2018MNRAS.479.2505K}.
The surface mass density is seen to  drop at R$_{25}$ (about 8.6~kpc), where the atomic disk warp becomes severe. 

To trace molecular hydrogen we use the IRAM 30 m all-disk $^{12}$CO J=2-1 survey of M33 presented in \citet{2014A&A...567A.118D}. The CO data cube has a spatial resolution of 12~arcsec and for the adopted distance of M33, this corresponds to a physical scale of 49~pc. The H$_2$ column density is computed using a CO-to-H$_2$ conversion factor X=N(H$_2$)/I$_{1-0}$=4$\times$10$^{20}$ cm$^{-2}$/(K~km~s$^{-1}$) \citep{2017A&A...600A..27G} and an intrinsic line ratio I$_{2-1}$/I$_{1-0}$=0.8 \citep{2014A&A...567A.118D}. All gas mass surface densities include the contribution chemical elements heavier than hydrogen.

The near-IR and mid-IR (NIR\ and MIR)\ emission of M33, from 3.6 to 70~$\mu$m, was investigated using data of the InfraRed Array Camera (IRAC) and Multiband Imaging Photometer for {\it Spitzer} (MIPS)  on board the {\it Spitzer} Space Telescope, and of the Wide-field Infrared Survey Explorer (WISE). 
The complete set of IRAC (3.6, 4.5, 5.8, and 8.0~\mi) and 
MIPS (24, and 70~\mi) images of \mm\ is described by \citet{2007A&A...476.1161V}.

For the far-IR (FIR), we made use of the Photodetector Array Camera and Spectrometer (PACS) and of the Spectral and Photometric Imaging Receiver (SPIRE) data on board of the Herschel Space Observatory mission. For the dust surface density map of M33, described in the next section, we used PACS data at 100 and 160 $\mu$m and SPIRE data at 250, 350, and 500 $\mu$m, from the Herschel M33 extended survey \citep{2010A&A...518L..67K,2010A&A...518L..69B}. Additional data at 70 $\mu$m were obtained from the Herschel program  OT2\_mboquien\_4 \citep{2015A&A...578A...8B}. For PACS, final maps were obtained directly from the Herschel Science Archive (HSA): they were provided as level 2.5 data and  processed with the JScanam map maker \citep{2017ASPC..512..379G}.
SPIRE lower level data were downloaded from the HSA and combined into maps using the
destriper module in the dedicated HIPE software \citep[][v15]{2011ascl.soft11001H}. 

\subsection{The stellar and dust surface density maps}

\begin{figure} 
\includegraphics [width=9 cm]{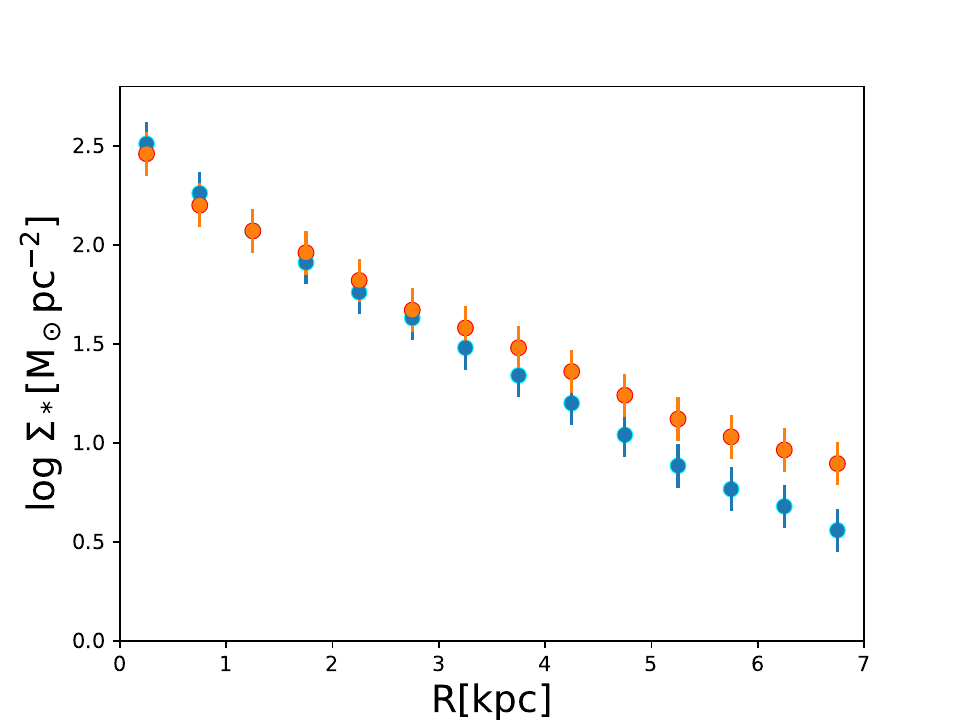}
\caption{ Face-on stellar mass surface density of M33 computed from the map of \citet{2014A&A...572A..23C} and averaged in radial bins for R<5~kpc  is shown as a function of galactocentric distance with cyan circles. Beyond 5~kpc  an extrapolation of the radial profile is shown. The orange filled circles indicate the average stellar mass density inferred  from the model of  \citet{2015ApJS..219....5Q}.}
\label{sigma}
\end{figure}

A map of the stellar mass surface density of M33 was derived by \citet{2014A&A...572A..23C} from multiband optical imaging using a fully Bayesian approach. It extends from the galaxy center to galactocentric distances of about 5~kpc. By extrapolating the radial profile of the stellar mass surface density out to the edge of the optical disk and adding an outer disk component, as observed by \citet{2011A&A...533A..91G}, the total stellar mass was estimated by  \citet{2014A&A...572A..23C} as 4.8$\times 10^9$~M$_\odot$.   Because of the lack of sensitivity of optical images at large galactocentric radii \citep{2014A&A...572A..23C}, we used the formula by \citet{2015ApJS..219....5Q} to estimate the stellar mass surface density out to the edge of the star-forming disk. This gives the stellar mass as a function of  galaxy distance and flux densities at 3.6 and 4.5~$\mu$m. In Figure~\ref{sigma}, we compared the mean stellar mass density  in radial bins using the formula of \citet{2015ApJS..219....5Q} with the results of  \citet{2014A&A...572A..23C}.  We can see that the combination of NIR fluxes
from the formula of \citet{2015ApJS..219....5Q} aptly approximate  the radial trend  of the stellar mass surface density traced by the map of \citet{2014A&A...572A..23C}; however, beyond 5~kpc, it predicts a shallower radial decline. This might be due to the lack of sensitivity of the optical stellar mass surface density beyond 5~kpc. The use of the NIR map implies an enclosed mass slightly higher than what the extrapolation of \citet{2014A&A...572A..23C} predicts (i.e. 4.8$\times 10^9$~M$_\odot$) without including the faint disk beyond 7~kpc. Given the similarities of the two radial profiles within 5~kpc, we  computed the local stellar mass surface density following the recipe of \citet{2015ApJS..219....5Q}  because the sensitivity of the NIR map allows us to trace the stellar mass density further out.

The dust surface density map was produced with the same procedure described in \citet{2017A&A...605A..18C}. In short, all Herschel maps have
been convolved to the lower resolution of the 500$\mu$m data and regridded on a common grid; FIR/submm emission templates based on the THEMIS dust model \citep{2017A&A...602A..46J} were fitted to the spectral energy distribution (SED) for $\lambda>100 \mu$m, for each pixel with significant S/N in all bands. For simplicity, it was assumed that the bulk of dust is heated by a single interstellar radiation field, scaled from the local one \citep{1983A&A...128..212M}. Because localised emission from dust heated by more intense interstellar radiation fields was not considered, the $70 \mu$m map is used as an upper limit to the fit \citep[see][for more details]{2017A&A...605A..18C}.
The dust mass surface density map obtained in this way has a resolution of a full width at half maximum of FWHM=36". To obtain a better resolution for making comparisons with the other maps, we assumed and tested that the ratio between the dust surface density and the 250 $\mu$m emission does not change significantly with the scale. This has allowed us to use the enhanced resolution HiRes $250 \mu$m map as a template to produce a dust surface density map with $FWHM\approx9"$.   The 250 $\mu$m map at a resolution that is higher than the nominal one (i.e., about half of the diffraction limited resolution of 18") has been produced using the HiRes procedure available within HIPE \citep{2014arXiv1401.2109X}.

\subsection{Disk dynamics and dark matter density}

The dark matter mass of M33 has been determined using dynamical analysis of its extended  rotation curve traced
via high resolution and sensitivity 21-cm imaging of the atomic gas \citep{2014A&A...572A..23C} complemented by the CO J=1-0 line emission in the inner regions.
The dynamical analysis shows the presence of a dark matter halo, as predicted by
numerical simulation of structure formation in a hierarchical $\Lambda$CDM universe. 
Although the rotation curve data extends out only to a maximum sampled radius,   the total virial mass of the galaxy has been recovered and it is lower (albeit non-negligible) compared to the M31 and MW halo masses.
We used the dark matter radial density profile from the best fitted Navarro-Frenk-White dark halo model given by  \citet{2014A&A...572A..23C}, with its parameters reported in Appendix~\ref{appsh},  to evaluate the contribution of the dark matter to the vertical gravitational force when computing the gas and stellar scale heights as a function of radius. We also used the rotation curve V(R) given by \citet{2014A&A...572A..23C}   to compute the epicyclic frequency or any other parameter connected  to the orbital velocity of the baryonic matter in   M33.

The stellar velocity dispersion, c$_s$, has been measured  in the central regions of M33 \citep{2007ApJ...669..315C} and  recently in a more extended area \citep{2022AJ....163..166Q}.  We used the mean stellar velocity  dispersion given by the more recent paper measured for galactocentric radii smaller than 5 kpc. There are no strong radial gradients, and the mean values going from the young to the intermediate and then to the old population increase slightly: from 15~km~s$^{-1}$ for the young population to 20~km~s$^{-1}$ for the old population. We used a constant value c$_s$ = 18~km~s$^{-1}$ in the main paper, considering the effects  of a  possible shallow radial gradient on the disk scale heights, suggested by the old stellar population in Appendix~\ref{appfuv}.

\subsection{Aperture photometry}

For the radial profiles, we considered coplanar concentric rings oriented in space as the galaxy disk; in projection, this implies using concentric elliptical apertures with axial ratios and position angle, as given by the M33 inclination and position angle. For individual regions, we performed circular aperture photometry along an hexagonal grid using maps at the original spatial resolution. This  is because the apertures we used are  larger than the spatial resolution of the maps. The lowest resolution map we used is the $^{12}$CO J=2-1 maps, with a FWHM of about 12~arcsec. 
We mostly used apertures with radii R$_a$=30~arcsec, although we tested our results, using  aperture radii of 20~arcsec,  or larger; in other words,  we sampled regions with minimum sizes of  40~arcsec, equivalent to 163~pc, to sizes of about 1~kpc.  
The 30~arcsec aperture is as wide as the largest giant molecular complexes found in M33 \citep{2018A&A...612A..51B}.
The hexagonal grid is ideal for fully sampling the maps. The side of the hexagon used is 1.5 $\times$ R$_a$  and we place the aperture in the center and vertices of the hexagons.

We masked out a few pixels in the HI dispersion map of M33, where the weakness of the HI signal did not allow us to measure the variance of $\sigma_{HI}$, the local 21-cm signal width. We also corrected the value of $\sigma_{HI}$ in the small areas where this is high ($\sigma_{HI}>25$~km~s$^{-1}$) due to local outflows or Milky Way contamination. We replaced the high $\sigma_{HI}$ with the corresponding radial average value. 

For $\Sigma_{dust}$ we masked out all pixels with undetermined dust mass surface density and in the analyses  involving the dust component, we included only the apertures for which less than  50$\%$ of the pixels have been masked out. For these apertures, the value of  $\Sigma_{dust}$ for the masked pixels has been set to zero. Excluding apertures where a large fraction of pixels have undetermined $\Sigma_{dust}$  implies  higher $\Sigma_{dust}$ mean values at large radii and, hence, a higher $\Sigma_{dust}/\Sigma_{gas}$ ratio  than what was computed using elliptical annuli (which include all the undetermined $\Sigma_{dust}$ pixels  set to zero).
We excluded these apertures only when the dust mass surface density was directly considered as one of the physical variables of our analysis.

\subsection{Star formation tracers}

Following the thorough review  by \citet{2012ARA&A..50..531K},  we  analyzed several star formation diagnostics, as given by extinction corrected emission at a particular wavelength or band, or by composite multiwavelength tracers. The fiducial approach used in our analysis is to derive  star formation over the past 10~Myr by combining the H$\alpha$ recombination line emission with 24$\mu$m emission following \citet{2007ApJ...666..870C} as

\begin{equation}
{SFR}[M_\odot\ yr^{-1} ] = 5.37\times 10^{-42} \Bigl( L_{H\alpha} + 0.031 L_{24\mu m} \Bigr)
,\end{equation}

\noindent
where the luminosities are given in erg~s$^{-1}$.
A galactic visual extinction value A$_v$= 0.114  was used towards M33 \citep{2011ApJ...737..103S}. A comparison of star formation surface densities inferred  using different tracers is presented in Appendix~\ref{appfuv}.

Corrections for IMF incompleteness become relevant when the total H$\alpha$ luminosity of the star-forming region is below 10$^{36}$~erg~s$^{-1}$ (see Sharma et al. 2011), which implies a star formation rate surface density below 0.1~M$\odot$~pc$^{-2}$~Gyr$^{-1}$ for an aperture radius of 30~arcsec and negligible 24$\mu$m emission. This condition was satisfied also in the faint sampled regions of the outer disk.

\subsection{Data  analysis}

When fitting straight lines to the data in the x-y plane, we used both a hierarchical Bayesian models (BAY) and orthogonal distance regressions (ODR). Both methods include uncertainties along the x- and the y- direction. For the ODR, which  minimizes orthogonal distances to the lines, we used ODR in the python package scipy.odr. We highlight the fact that  using ODR without accounting for data  uncertainties gives different estimates of the fitted line parameters. Since we are sampling the whole star-forming disk of M33, uncertainties are necessary in order to include faint regions in the outer disk as well as interarm regions.  For the Bayesian modeling, we used the LinMix package (Kelly 2007). We give a detailed description of the uncertainties of physical quantities involved in scaling relations in Appendix~\ref{apperr}.

Although scaling relations are  used to understand the physical processes that drive the formation of stars,  correlations do not imply causality. To rank the importance of variables and to find the driver for the star formation process (or what is typically referred to as the fundamental relations), we used a random forest (RF) approach. This type of analysis is also powerful for extracting non-linear dependencies in the data. A RF consists of a series of decision trees used to asses the relative importance of input variables in determining a target variable. Our approach follows the analyses presented by \citet{2020MNRAS.493L..39E},
\citet{2022MNRAS.510.3622B},
\citet{2023MNRAS.519.1149B}, and
\citet{2024MNRAS.52710201E} (see also \citealt{2022A&A...659A.160B} for more details).

\section{Disk thickness and midplane pressure}
\label{secphy}

We computed the midplane pressure and the vertical scale heights of gas and stars in the M33 disk as function of galactocentric distance using equilibrium conditions and plane parallel approximation. The matter vertical distribution is regulated by the local gravitational potential due to stars, gas, and dark matter and by the velocity dispersions of the two baryonic components, the gas, and the stars. We describe in Appendix~\ref{appsh} the details of our numerical computation and the data used to find the local pressure, gas, and stellar scale heights. 

Due to limited data availability and/or for simplicity, it has become common practice to adopt a simplified analytic approximations for the midplane pressure. We compared the resulting numerical values of the midplane pressure to those inferred from analytic approximations, often used in the literature. These are called the hydrostatic pressure, P$_{hy}$, as defined by \citet{1989ApJ...338..178E}, the pressure P$_{BR}$, as given by \citet{2004ApJ...612L..29B}, and the dynamical equilibrium pressure, P$_{DE}$, as expressed by \citet{2011ApJ...743...25K,2022ApJ...936..137O}.  We  explicit all these approximations for the midplane pressure,  considering or neglecting dark matter in the disk,  and several estimates of the gas and stellar scale heights in Appendix~\ref{appsh}. 
The comparison between the computational values and those derived from  analytic expressions (shown in Appendix~\ref{appsh}) suggest to use P$_{hy}^{1d}$ as the best proxy for the midplane pressure across the star-forming disk of M33, which is expressed as 

\begin{equation}
P_{hy}=\rho_g c_g^2  \simeq {\pi G \over 2} \Sigma_g \Bigl\lbrack\Sigma_g +\Sigma_s^g +\Sigma_{dm}^g\Bigr\rbrack= {\pi G \over 2} \Sigma_g \Bigl\lbrack \Sigma_g + {c_g\over c_s} \Sigma_s  + 2 \rho_{dm} h_g \Bigr\rbrack
.\end{equation}

\noindent
This can be re-written using the expression for the gas scale height h$_g$, derived in Appendix~\ref{appsh}, as

\begin{equation}
P_{hy}={\pi G\Sigma^2_g \over 4}  \Biggl\lbrace 1+{ c_g\Sigma_s \over  c_s\Sigma_g} + \sqrt{\Bigl\lbrack 1+{ c_g\Sigma_s \over c_s\Sigma_g}\Bigr\rbrack^2+8 {c_g^2 \rho_{dm}\over \pi G \Sigma_g^2}  }\  \Biggr\rbrace
.\end{equation}

\noindent
In the above expressions G is the gravitational constant, and all other variables are function of the galactocentric radius, R. The symbols  $\Sigma_g$ and $\Sigma_s$ indicate the total mass surface density of gas and stars perpendicular to the galactic plane, while $\Sigma_s^g$ and $\Sigma^g_{dm}$ are the stellar and dark matter column densities within the gas layer, respectively. The gas mass density in the midplane is $\rho_g$  and c$_g$ is the gas velocity dispersion perpendicular to the plane, which includes the effect of nonthermal pressure components due to cosmic rays and magnetic fields. Following \citet{2019ApJ...882....5W} and taking into account that M33 is dominated by atomic gas and has similar values of CO and HI dispersions, we used $c_g=\sqrt{1.3} c_g^0 =\sqrt{1.3} \sigma_{HI}$, with $\sigma_{HI}$ the atomic gas velocity dispersion along the line of sight. The above formula is an extension of the approximation suggested by \citet{1989ApJ...338..178E} because it includes dark matter. We  assumed that the 
stellar vertical extent is larger than the gas vertical extent (assumption verified by our results). As discussed in the Appendix~\ref{appsh}, the dark matter density in the above expression is $\rho_{dm}=1.32 \rho_{dm}^0$, where $\rho_{dm}^0$ is the dark matter density at midplane and $\Sigma_{dm}^g=2 \rho_{dm} h_g$, with h$_g$ as the gas scale height given explicitly in Appendix~\ref{appsh}.

\noindent
To investigate star formation drivers we also test if  simplified  pressure expressions are reliable predictors of the star formation rate surface density $\Sigma_{SFR}$. For this purpose, we selected the expression given by \citet{2004ApJ...612L..29B,2006ApJ...650..933B}, labeled P$_{BR}^0$ in Appendix~\ref{appsh}, which has been used by several authors \citep[e.g.][]{2020ApJ...892..148S,2021MNRAS.503.3643B,2024MNRAS.52710201E}. Authors often refer to P$_{BR}^0$ as the dynamical equilibrium pressure, P$_{DE}$, because it has been quoted by OK22 as the limiting form of their more complete expression for the pressure. However, as underlined in Appendix~\ref{appsh}, the full analytic expression of P$_{DE}$  (as derived by OK22) is different than P$_{BR}^0$, as pointed out by the authors themselves. To use the same notation as other works, we defined P$_{DE}^*\equiv$ P$_{BR}^0$, neglecting dark matter and nonthermal components and under the assumption of a radially constant vertical scale height for the stellar distribution. We refer to this as the simplified dynamical equilibrium pressure which can be written as  
\begin{equation}
P^*_{DE} \equiv P_{BR}^0 = {\pi \over 2} G \Sigma^2_g +\Sigma_g c^0_g \sqrt{2G  \rho_{s}},  \qquad {\hbox{with\ }} \rho_s\simeq{7.3 \Sigma_s\over 4 l_s }
,\end{equation}

\noindent
where l$_s$ is the stellar radial scale length \citep{2008AJ....136.2782L}.  We underline that under the assumptions of no dark matter the stellar surface density needs to be much greater than the gas surface density for P$_{DE}^*$ to be equivalent to the more generic expression of P$_{DE}$ (reported in Appendix~\ref{appsh}). This is never the case for the M33 disk. 

\begin{figure} 
\includegraphics [width=9 cm]{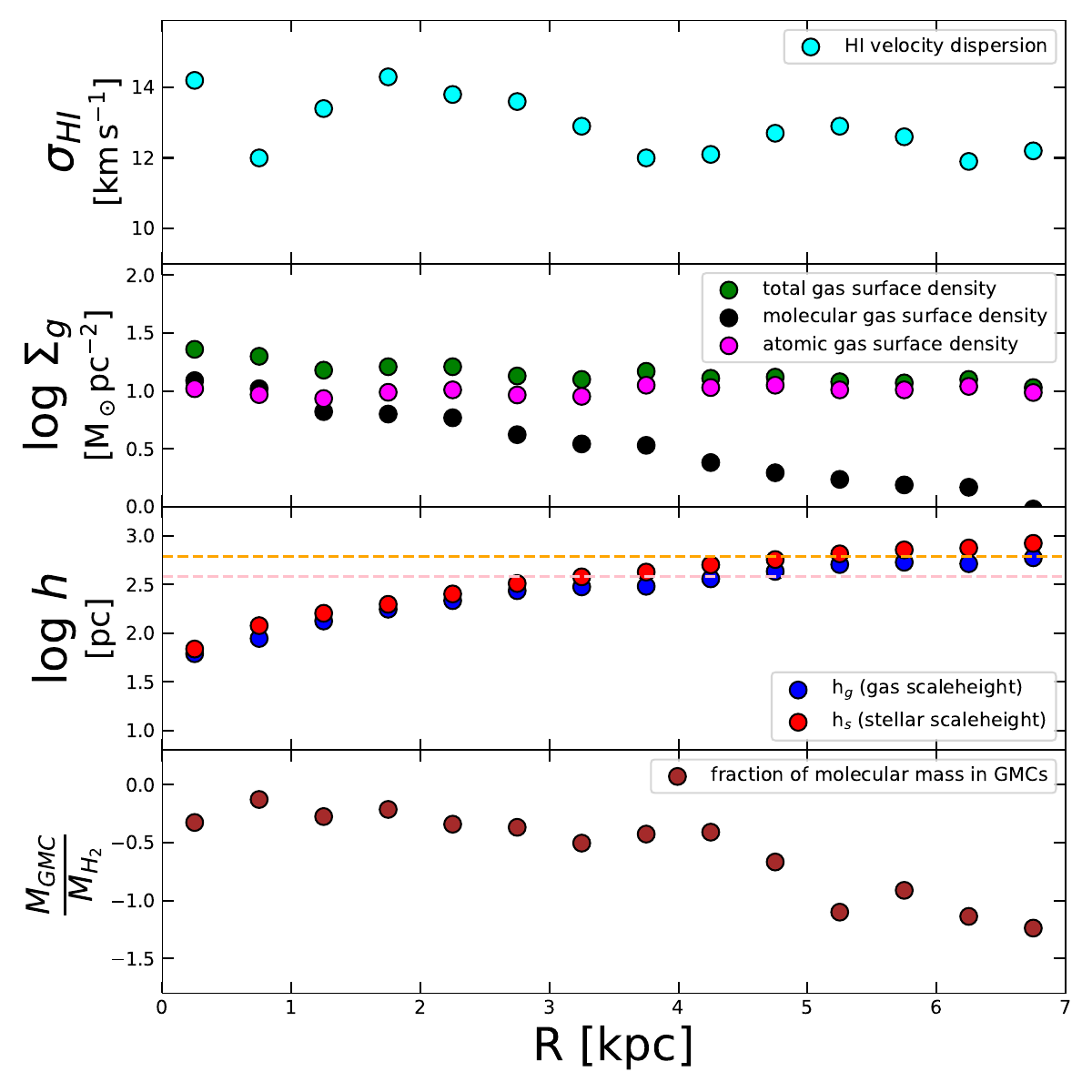}
\caption{
Radial averages in tilted concentric rings, oriented spatially as the M33 disk, are shown as a function of  the ring mean galactocentric distance, R.  From top to bottom, we show: Mean HI velocity dispersion along the line of sight, the gas surface densities (total, molecular, and neutral atomic gas corrected for helium) perpendicular to the M33 disk, the estimated gas, and stellar scale heights perpendicular to the galactic plane, the fraction of molecular mass locked into GMCs. The dashed pink and orange lines in the scale height panel show the constant stellar scale height often used related to the radial stellar scale length and to the half-light radius respectively (see text for details).}
\label{scale_r}
\end{figure}

In Figure~\ref{scale_r}, we show the mean HI velocity dispersion, the atomic, molecular and total gas mass surface densities, and the gas and stellar scale heights (given explicitly in Appendix~\ref{appsh}) as functions of galactocentric distance.
These have been computed using photometric measurements in elliptical annuli. In Figure~\ref{scale_r}, we can see that the average HI velocity dispersion is slightly lower than the average stellar velocity 
 dispersion (18~km~s$^{-1}$), with its mean values which vary between 14~km~s$^{-1}$ to 12~km~s$^{-1}$. The modest radial variation  of $\sigma_{HI}$ suggests the ubiquitous presence of turbulence at all radii although the small width of the probability density function indicates a decrease in the Mach number at large galactocentric radii \citep{2018A&A...617A.125C}.  Figure~\ref{scale_r} also shows that the gas in M33 is mostly in the atomic phase with the molecular gas surface density slightly above the atomic gas surface density only in the innermost kpc. The radially steady decrease in the molecular fraction  in M33 is likely due to changes in the local characteristics of the ISM, such as the pressure and gas volume density, given the almost constant gas surface density in the star-forming disk.

The dashed lines in the scale height panel indicate the constant scale height values h$_{s,0}$=2$\times$ l$_s$/7.3 and h$_{hrl}$=2 R$_{hlr}$/($1.68\times7.3$), shown as pink and orange dashed lines, often used  as a proxy to h$_s$ in nearby galaxies \citep[e.g.][]{2008AJ....136.2782L,2024MNRAS.52710201E}. The l$_s$ and R$_{hlr}$ are the stellar radial scale length and the half-light radius which  for M33 are l$_s$ = 1.4~kpc, R$_{hlr}$=3.8~kpc  \citep{1994ApJ...434..536R,2001A&A...369..421B}. We underline that our scale height, defined as h=$\Sigma/2\rho$, where $\rho$ is the midplane density and $\Sigma$ the total surface density of matter perpendicular to the disk,  would be twice the exponential scale height if we were to consider the exponential behavior of  an  isothermal self-gravitating disk  at large distances from the midplane.  Because this exponential height is usually measured for other galaxies, we have   $\Sigma/4\rho$= l$_s$/7.3 (see Appendix~\ref{appsh} for more details). In Figure ~\ref{scale_r}, we show  also the HI velocity dispersion in M33 along the line of sight and radial averages of the face-on values of HI, H$_2$ and total gas surface densities.  In the bottom panel of Figure~\ref{scale_r}, we show the fraction of the molecular mass which forms giant molecular clouds (GMCs), as catalogued by \citet{2017A&A...601A.146C}. The spatial resolution of the CO map used for the GMC catalogue is 49~pc and the completeness limit around 6$\times 10^4$~M$_\odot$.  The fraction of molecular mass locked into GMCs decreases significantly beyond 4~kpc, from about 40$\%$ to about 10$\%$. This can be due to an effective increase in the diffuse molecular gas fraction or to the presence of low mass clouds, with radius below the spatial resolution of the survey used, which become more pervasive in the outer disk. Although the cloud mass spectra steepens as we move from the inner to the outer disk \citep{2018A&A...612A..51B}, we cannot quantify the amount of gravitationally bound clouds versus a more pervasive, truly diffuse molecular gas. Additional radial profiles are shown in Appendix~\ref{apprad}.

\section{Star formation drivers across the M33 disk}

In this section, we examine the possible drivers of the star formation rate surface density, $\Sigma_{SFR}$, across the M33 disk. We examined the whole disk, sampling it with circular apertures of 30~arcsec in radius (244~pc wide apertures) and also the  inner and outer star-forming disk separately. In addition, we removed radial dependencies to investigate local azimuthal trends and test correlations at different spatial scales, from 160~pc to about 1~kpc. Variations of some of our working assumptions, such as the radial gradient of the stellar velocity dispersion and of the star formation tracer, are discussed in Appendix~\ref{appfuv}.

\subsection{Relating star formation rate surface density to other physical quantities}

\begin{table*}
\caption{Scaling relations from orthogonal distance regression (ODR) and hierarchical Bayesian method (BAY) - R$_a$=30''}            
\label{table:1}       
\centering                                      
\begin{tabular}{c c c cc c c  c c }          
\hline\hline                        
Relation & $\alpha_{ODR}$ & $\beta_{ODR}$ & $\alpha_{BAY}$ & $\beta_{BAY}$ & R$^2_{ODR}$ & C$_{BAY}$ & S$^O_{BAY}$  &  S$^I_{BAY}$   \\     
\hline                                   
 log$\Sigma_{SFR}$--log$\Sigma_{H_2}$(All)   & 1.48$\pm$0.02  &-0.45$\pm$0.01   & 1.37$\pm$0.03 & -0.43$\pm$0.02    & 0.43 &0.70 & 0.32 & 0.21\\       
 log$\Sigma_{SFR}$--log$\Sigma_{H_2}$(In)    & 1.24$\pm$0.04 & -0.28$\pm$0.03 & 1.13$\pm$0.04  & -0.24$\pm$0.03     &0.45  &0.69 & 0.28 & 0.20  \\     
 log$\Sigma_{SFR}$--log$\Sigma_{H_2}$(Out)  & 2.17$\pm$0.08 & -0.65$\pm$0.02 & 2.67$\pm$0.12  & -0.45$\pm$0.03    &-0.63  &0.75 & 0.49 & 0.06  \\      
 \hline
 log$\Sigma_{SFR}$--log$\Sigma_{*}$(All)         & 1.08$\pm$0.02  &-1.34$\pm$0.03   & 0.97$\pm$0.03 & -1.23$\pm$0.03    & 0.45 &0.69  & 0.32 & 0.23 \\     
 log$\Sigma_{SFR}$--log$\Sigma_{*}$(In)          & 1.24$\pm$0.04 & -1.67$\pm$0.07 & 1.15$\pm$0.04  & -1.54$\pm$0.08     &0.47  &0.66 & 0.28 & 0.22  \\      
 log$\Sigma_{SFR}$--log$\Sigma_{*}$(Out)       & 1.33$\pm$0.05  & -1.59$\pm$0.06 & 0.82$\pm$0.07  & -1.04$\pm$0.07    &-0.28  &0.49 & 0.33 & 0.25  \\      
 \hline
 log$\Sigma_{SFR}$--logP$_{hy}$/k$_B$(All)    & 1.26$\pm$0.02  &-5.31$\pm$0.08   & 1.20$\pm$0.02 & -5.03$\pm$0.09    & 0.61 &0.72  & 0.27 & 0.18  \\      
 log$\Sigma_{SFR}$--logP$_{hy}$/k$_B$(In)     & 1.05$\pm$0.03  & -4.32$\pm$0.12 & 1.01$\pm$0.03  & -4.13$\pm$0.14    &0.62  &0.72 & 0.24 & 0.16  \\      
 log$\Sigma_{SFR}$--logP$_{hy}$/k$_B$(Out)  & 1.65$\pm$0.05  & -6.93$\pm$0.20 & 1.42$\pm$0.06  & -5.99$\pm$0.25    &0.18  &0.66 & 0.28 & 0.18  \\      
 \hline
 log$\Sigma_{H_2}$--logP$_{hy}$/k$_B$(All)     & 0.84$\pm$0.01  &-3.20$\pm$0.05   & 0.80$\pm$0.02 & -3.02$\pm$0.05    & 0.74 &0.72  & 0.16 & 0.13 \\       
 log$\Sigma_{H_2}$--logP$_{hy}$/k$_B$(In)      & 0.83$\pm$0.02  &-3.16$\pm$0.08   & 0.79$\pm$0.02 & -2.99$\pm$0.08    & 0.71 &0.74  & 0.16 & 0.10  \\       
 log$\Sigma_{H_2}$--logP$_{hy}$/k$_B$(Out)    &0.77$\pm$0.02  &-2.92$\pm$0.10   & 0.76$\pm$0.03 & -2.89$\pm$0.11     & 0.31 &0.74  & 0.16 & 0.04  \\      
 \hline                                             
\end{tabular}
\end{table*}

\begin{figure*} 
\includegraphics [width=12.8 cm]{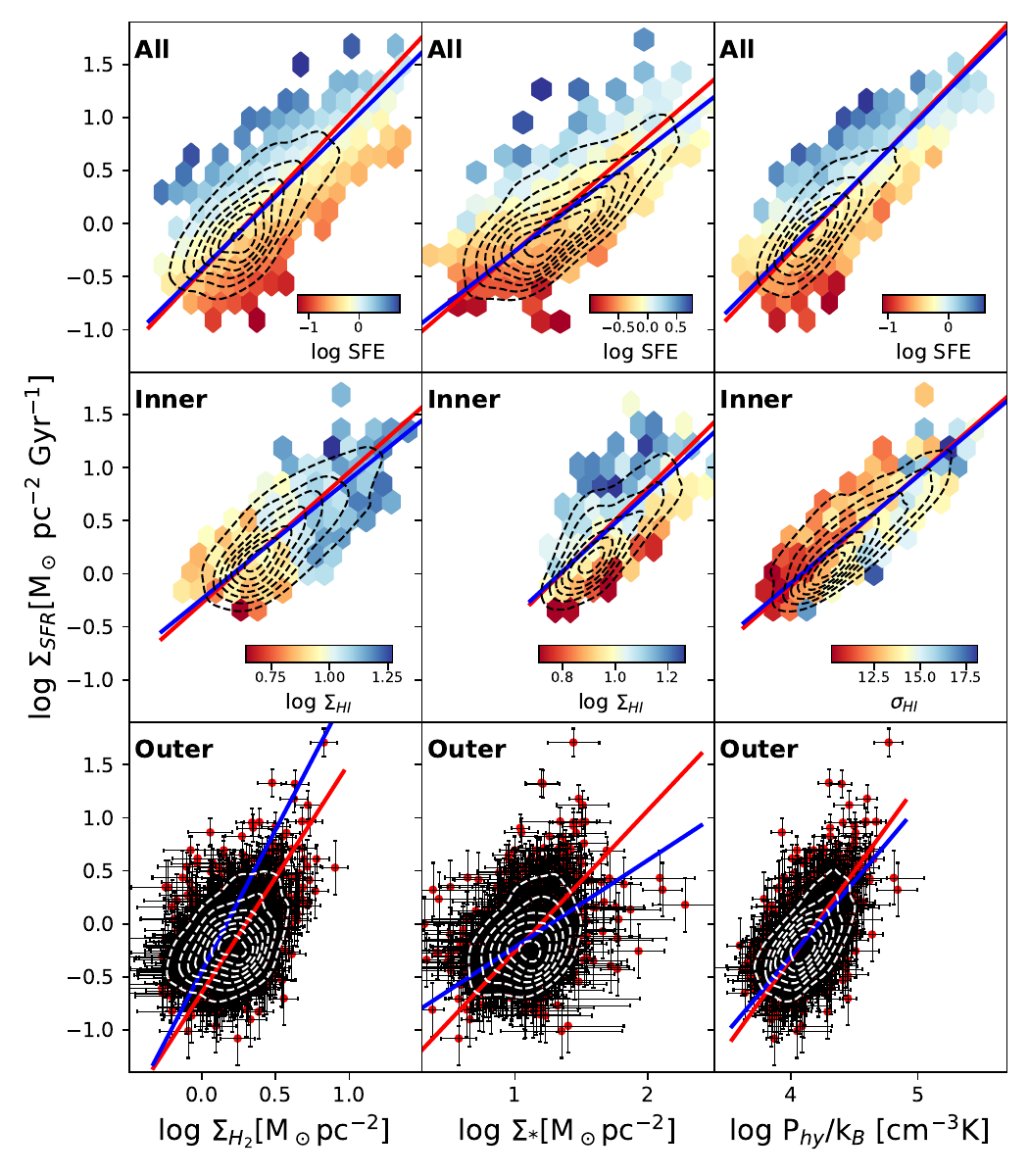}
\includegraphics [width=5.5 cm]{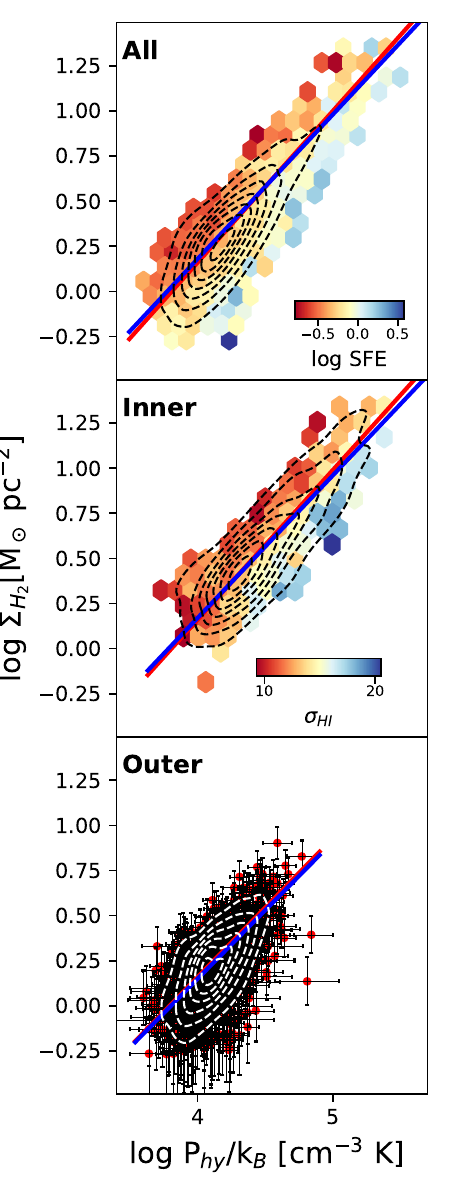}
\caption{Data and the relative linear relations, plotted in red for ODR and in blue for BAY method.  The top panels refer to the whole sample, where color coding is according to the log of the effective SFE, readable in the color bar. Data and fits in the  middle panels refer to the inner disk and are color coded with a third variable as indicated by the color bar: either log $\Sigma_{HI}$ in M$_\odot$~pc$^{-2}$ or $\sigma_{HI}$ in km~s$^{-1}$. In the bottom panel, we display the data relative to the outer disk with the relative rms. In all panels, the dashed contours are drawn according to the density of data in the plane.}
\label{pan9}
\end{figure*}

In Table~1, we display the slopes $\alpha$ and intercepts $\beta$ of the fitted linear relations for the logarithm of the following physical quantities: $\Sigma_{SFR}$--$\Sigma_{H_2}$ (KS law), $\Sigma_{SFR}$--$\Sigma_{*}$ (SFMS), $\Sigma_{SFR}$--P$_{hy}$/k$_B$, and $\Sigma_{H_2}$--P$_{hy}$/k$_B$. Units of  $\Sigma_{SFR}$ are [M$_\odot$pc$^{-2}$Gyr$^{-1}$], mass surface densities are in [M$_\odot$pc$^{-2}$], and pressure is in [cm$^{-3}$K]. We show the resulting fits using both the ODR and the BAY method with the uncertainties listed in Appendix~\ref{apperr}.The degree of linear correlation is quantified using the R$^2$ test for ODR fits and the latent variable correlation, C, for hierarchical Bayesian fits. For this last method, we also show the observed and intrinsic scatter indicated by the symbols S$_{BAY}^O$ and S$_{BAY}^I$, respectively. All these parameters refer to the whole sample for R$\le 7$~kpc (label: All) or to the inner and outer disk separately (label In for R$\le 4 $~kpc, and Out for 4$<$R$\le 7$~kpc). The data and the relative linear relations, plotted in red for ODR and in blue for BAY method, are shown in Figure~\ref{pan9}.  The data for the whole and inner disk are color coded according to a third variable, which is possibly related to the scatter along the y-axis as indicated by the color bar. In the top panels, for the whole sample and for all the relations, we used the effective star formation efficiency (SFE) in Gyr$^{-1}$. In the middle panels for the KS and SFMS, the color coding is done with  log $\Sigma_{HI}$, which is related to the scatter along the y-axes for the SFMS. Relations shown in the two rightmost panels of the middle row  have been color-coded with $\sigma_{HI}$ in km~s$^{-1}$. In the bottom panel, we display the data relative to the outer disk with the relative rms. In all panels, the dashed contours are drawn according to the density of data in the plane.

The ODR and BAY methods give slopes of the scaling relations that are consistent with each other, namely, within three times the quoted uncertainties, for both the global and inner disk data. Some discrepant results are found for the outer disk when correlations are not strong as indicated by the R$^2$ test, such as those for the KS and SFMS outer disk relations. Slopes retrieved by BAY fits are generally a bit steeper than those retrieved by the ODR fits.

The strongest correlation, as given by the R$^2$ and by the lowest scatter, S$^O_{BAY}$ and S$^I_{BAY}$, was found for the molecular mass surface density as a function of the local pressure, the log P$_{hy}$--log $\Sigma_{H_2}$ relation, shown in the rightmost panels of Figure~\ref{pan9}. Pressure is the key ingredient to form molecules enhancing the ISM density.  Because its relation with $\Sigma_{H_2}$ is even tighter than that established by $\rho_g$ alone, turbulence motions  clearly plays a contributing role  in controlling H$_2$ by pressure. The sublinear slope of the relation, consistent between the inner and outer disk, hints to other variables involved in the atomic to molecular gas conversion, such as the dissociating interstellar radiation field. 

With respect to  the relations with $\Sigma_{SFR}$, the variable which established the tightest  relation, as given by the R$^2$ and by the lowest scatter, is P$_{hy}$.
The scatters of the resolved KS law and the resolved SFMS in M33 are larger than that of the log $\Sigma_{SFR}$-log $P_{hy}$ relation. 
The log $\Sigma_{SFR}$-log $P_{hy}$ relation has a slope that is compatible with a linear scaling of the $\Sigma_{SFR}$ with P$_{hy}$. Some correlation is found also in the outer disk although the slope of the relation is  steeper, close to 1.5. The KS relations have superlinear slopes, with a large scatter in the outer disk,  where the correlation is almost lost, as can be inferred by the negative R$^2$ value. 

Effective SFE variations drive the dispersion along the y-axis for all the relations examined and across the whole star-forming disk (see top panels of Figure~\ref{pan9}). For the KS law (top-left panel in Figure~\ref{pan9}), the scatter is related to SFE as expected: at a given molecular gas surface density, regions with a higher $\Sigma_{SFR}$ are also those of higher efficiency for star formation. These variations hold also for a given value of P$_{hy}$ (top row, third panel from the left) because of the tight relation between P$_{hy}$ and $\Sigma_{H_2}$.

\begin{figure} 
\centering
\includegraphics [width=9.0 cm]{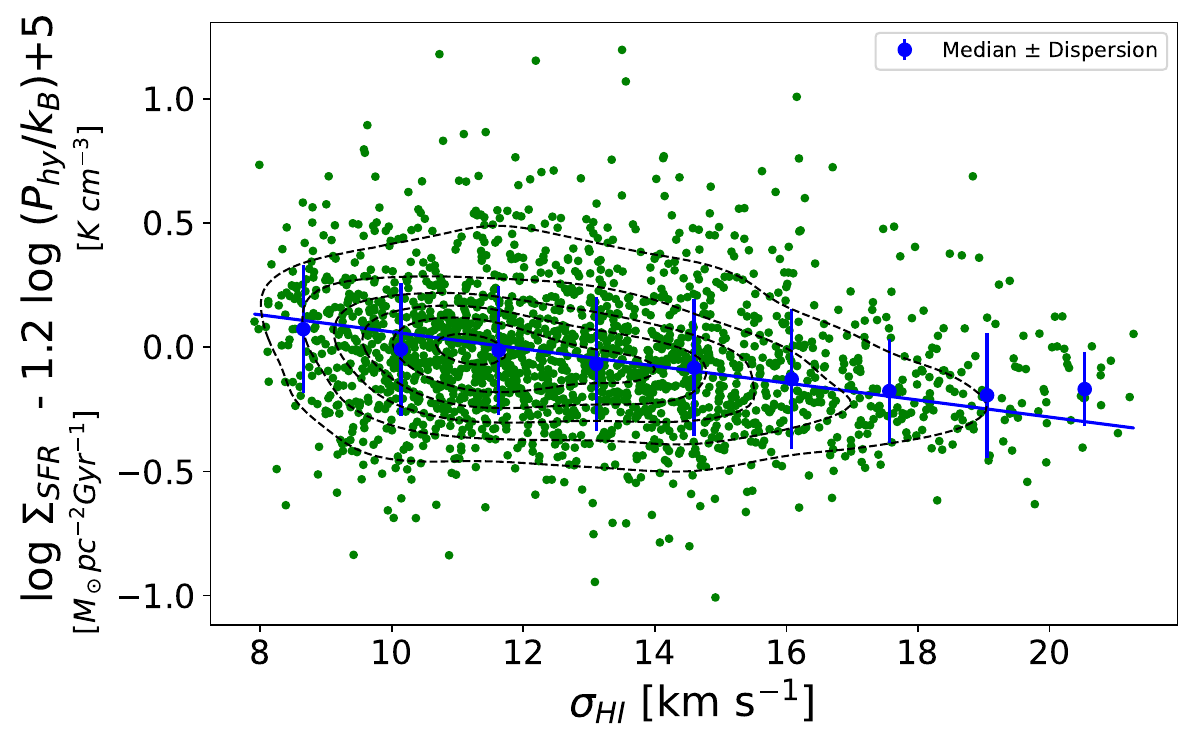}
\caption{ Residual correlation between  $\Sigma_{SFR}$,  and  the gas dispersion, $\sigma_{HI}$, after removing  the dependence of $\Sigma_{SFR}$ from the local hydrostatic pressure. Dashed contours indicate data density in the plane.}
\label{sigmaHI}
\end{figure}

In addition, we note that for a given P$_{hy}$ in the inner disk, the gas velocity dispersion, $\sigma_{HI}$, is low when $\Sigma_{SFR}$ is high (third column, second row) because at low $\sigma_{HI}$, the midplane density is high and the molecular hydrogen mass surface density increases (fourth column, second row).  This decrease in $\Sigma_{SFR}$ with $\sigma_{HI}$ was also found in \citep{2022ApJ...928..143E}. We explicit this residual correlation between $\Sigma_{SFR}$ and $\sigma_{HI}$ by fitting $\sigma_{HI}$ versus the residual around the mean of the $\Sigma_{SFR}-P_{hy}/k_B$ correlation, given by the BAY method. The fit, shown in Figure~\ref{sigmaHI},  gives a slope  of $-0.034$ and intercept $0.405$, which combine to give the three-parameter correlation:

\begin{equation}
{\hbox{log}}\Sigma_{SFR} = 1.20\  {\hbox{log}} \ \Bigl({\frac{P_{hy}}{k_B}}\Bigr) - 0.034\  \sigma_{HI} - 4.6
.\end{equation}

For the SFMS (second column of Figure~\ref{pan9}) the scatter  perpendicular to the sequence is related to the HI content of the regions, with HI-rich regions showing a higher SFR compared to HI-deficient regions. 
The enhancement of the SFR in HI rich regions  at a given radius, as we shall see later, is due to the fact that it is the pressure, a combination of gas and stellar surface density, which drives the star formation and not the stellar mass density alone. Moreover, because $\Sigma_{HI}$ is not as directly related to $\Sigma_{SFR}$ as $\Sigma_{H_2}$,  the $\Sigma_{HI}$ gradient is orthogonal to the sequence. In the outer disk, where the atomic gas surface density drives pressure variations, there is no SFMS correlation.

The relation between log $\Sigma_{SFR}$ and  log P$_{hy}$ is  not a single power law because there is  a flattening for log P$_{hy}/k_B > 4.7$~cm$^{-3}$~K and log $\Sigma_{SFR}>0.5$~M$_\odot$pc$^{-2}$Gyr$^{-1}$. This pressure value for the transition to the flatter distribution  is similar to the value of P$_{DE}/k_B$ for which \citet{2024MNRAS.52710201E} also find a flattening in the distribution of $\Sigma_{SFR}$ for the extended ALMaQUEST sample. Although  the flattening is driven by starburst galaxies in the  ALMaQUEST sample, this feature is also present for their control sample (see their Fig. 4). The $\Sigma_{SFR}$ at transition is however lower by 0.5~dex than the value found by \citet{2024MNRAS.52710201E} using extinction corrected  H$\alpha$ line emission to trace $\Sigma_{SFR}$.  It should be underlined that M33 is at the low end of the stellar and molecular mass distribution of the sample  considered by \citet{2024MNRAS.52710201E} and their sample is not representative of low luminosity blue galaxies such as M33.

The change of slope for the  log $\Sigma_{SFR}$ -- log P$_{hy}$ relation becomes more evident when plotting this relation  separately for  the inner and  outer disk, as we show in the middle and lower panels of Figure~\ref{pan9}.  The pressure range for the outer disk is smaller,  the uncertainties  are higher, and the scatter increases, which  all make the slopes more uncertain than for the inner disk.  The density contours of the data distribution also indicate a steeper decrease in $\Sigma_{SFR}$  with decreasing $P_{hy}/k_B$  toward the outer disk  compared to the inner disk.  This faster decrease in $\Sigma_{SFR}$ cannot be driven by a faster decrease  in the available fuel for star formation in the outer regions because the relation between $\Sigma_{H_2}$ and P$_{hy}$ stays the same across the whole star-forming disk (rightmost column). It is possible that it is the SFE per unit mass of molecular gas, which decreases faster with pressure in the outer disk than the inner disk.  This may be caused by an increase in the diffuse fraction of molecular hydrogen  toward the outer disk,  which is suggested also by the decreasing ratio of GMC mass to total gas mass in Fig. \ref{scale_r}. Diffuse molecular gas can still be dependent on pressure, but it does not collapse directly into star-forming clouds like GMCs do. Such a change in the diffuse molecular fraction might result from the lack of spiral arms, which collect small diffuse clouds into large cloud  complexes that are able to cool and collapse more efficiently. The outer disk decline of the SFE is not alleviated if the CO-to-H$_2$ conversion factor increases there.

\subsection{The star formation timescale}

\begin{figure} 
\hspace{-0.5 cm}
\includegraphics [width=10. cm]{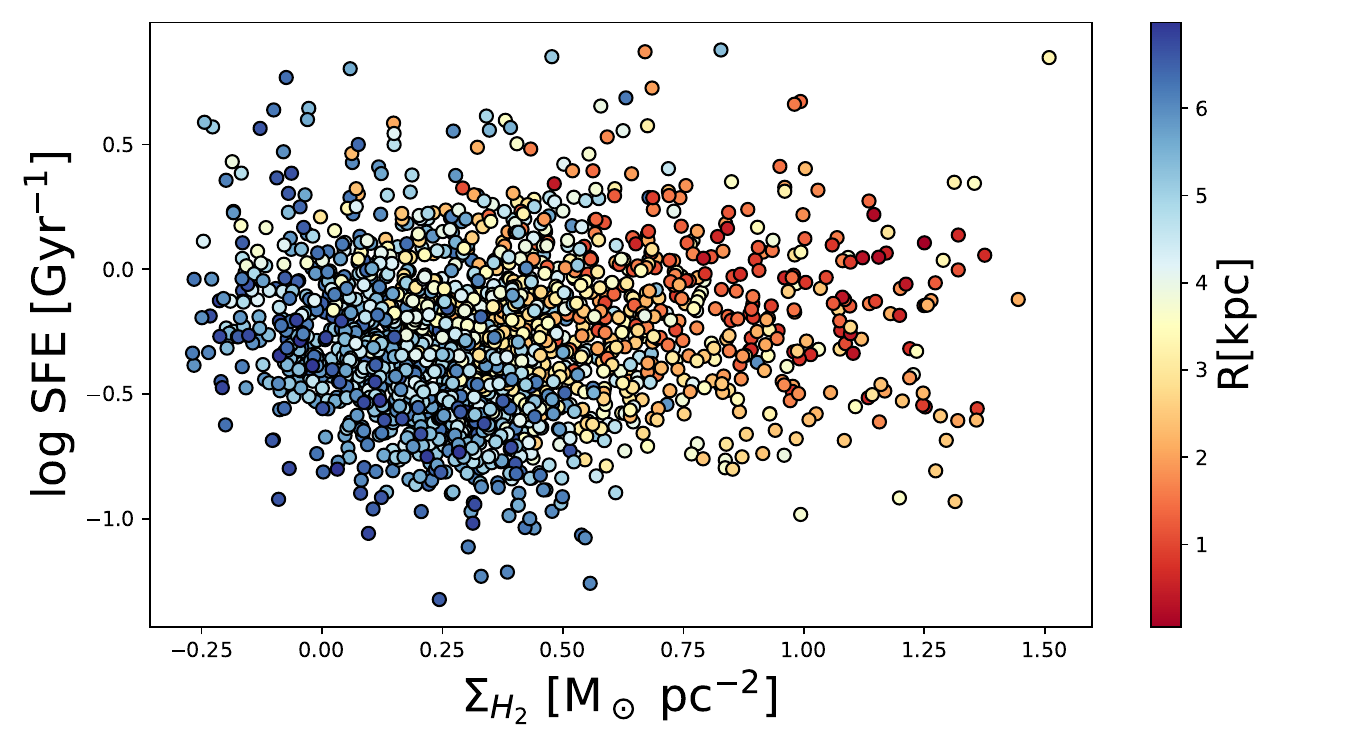}
\caption{Efficiency of star formation SFE=$\Sigma_{SFR}/\Sigma_{H_2}$ as a function of the molecular mass surface density, $\Sigma_{H_2}$, for regions at various galactocentric radii, as indicated by the color bar.}
\label{age}
\end{figure}

The scatter in the effective SFE ($\Sigma_{\rm SFR}/\Sigma_{\rm H_2}$ in Figure \ref{pan9}) should contain some contribution from the stochasticity of physical processes during gas collapse to star formation \citep[e.g.,][]{2016ApJ...833..229L}. However, on the small spatial scales considered here, it can also have a contribution from the evolution of the star formation process and of the hosting cloud -- a factor that rarely needs consideration \citep{2010ApJ...722.1699S,2017A&A...601A.146C}. This is because the H$\alpha$+24$\mu$m luminosity does not actually measure the SFR, but only the current luminosity of stars. In a large region with a fully sampled IMF and enough GMCs to sample all the stages in their life-cycle, this luminosity can be divided by a representative timescale, such as the lifetime of a massive star in the case of H$\alpha$, to give a good measure of the average SFR. However, in a small region or for particular GMCs, the luminosity should really be viewed as the luminosity of the stars that have formed up to now. A particular cloud that is just beginning to form stars will have a low H$\alpha$+24$\mu$m or other young-star-related luminosity for its molecular mass, while a similar GMC that has been forming stars for a longer time will have a greater number of stars today and a higher young-star luminosity for its molecular mass. If we consider that the first appearance of  star-formation tracers in our sample occurs after $\sim0.5$ Myr in a GMC, and the last stars form at GMC break-up, then the ratio of present-day young-star luminosity to present-day molecular mass can vary  by a factor of a few for a constant SFR in the cloud.   A similar effect has been found in high spatial resolution studies of the M33 molecular cloud lifecycle  \citep{2010ApJ...722.1699S}. The break-up timescale may be slightly longer than the duration of the embedded phase, which for M33 molecular complexes, comparable in size to the sampled regions, is on the order of 2.5~Myr \citep{2017A&A...601A.146C}. In fact the youngest stellar clusters at breakup have estimated ages of 3.5~Myr \citep{2017A&A...601A.146C}.

To illustrate this possibility we show in Figure~\ref{age} the log of the SFE as a function of log $\Sigma_{H_2}$. As star formation proceeds  for 3 to 4 Myr, the total luminosity of the young stars increases and the molecular mass decreases by a small amount, causing the region to move mostly upward in the plot. We  color-coded the points with the galactocentric radius to distinguish this evolutionary effect from  SFE variations that depend on radius. We can see in Figure~\ref{age} that the SFE varies by about a factor of 3 at a given galactocentric distance. 
We also observed a decrease in the ratio of the two terms in $\Sigma_{SFR}$, the 24$\mu$m to H$\alpha$ ratio, as we move along the positive y-axis in Figure~\ref{age}, due to cluster and molecular cloud evolution as young stars break through the cloud. This variation in the effective SFE due to the duration of the star formation process could even dominate the observed scatter in the KS law and $\Sigma_{SFR}$--P$_{hy}$ relation at our spatial resolution. For this reason, we color-coded with the effective SFE  data in Figs.~\ref{pan9} and~\ref{age}.  The SFE scatter from GMC evolution should be smaller for less sensitive observations, where the first stars become visible later, and it would also be smaller  for larger regions that contain several GMCs at different evolutionary stages (see  Fig.~\ref{aper}).

\subsection{Radially normalized quantities}

 To remove the average radial trends, we examined the scaling relations using fluctuations in the azimuthal direction \citep[see][]{2024ApJ...966..233E}.  We do this  by subtraction in log scale, which implies dividing physical quantities by their median radial values. We call these quantities R-normalized quantities. 
By searching for possible correlations between R-normalized log $\Sigma_{SFR}$ and other R-normalized physical quantities, we find that R-normalized $\Sigma_{H_2}$, $\Sigma_{HI}$, and P$_{hy}$ correlate  with R-normalized $\Sigma_{SFR}$ well. Furthermore, there is no leftover correlation between $\Sigma_*$ and $\Sigma_{SFR}$.
Pressure P$_{hy}$ is the most significant variable that has good R-normalized scalings. Other than with $\Sigma_{SFR}$, P$_{hy}$ also exhibits a well-defined correlation with the gas (either total, molecular, or atomic), although the significance of the correlation is small with  R$^2$ on the order of 0.5. 
  
\begin{figure*} 
\centering
\includegraphics [width=18.0 cm]{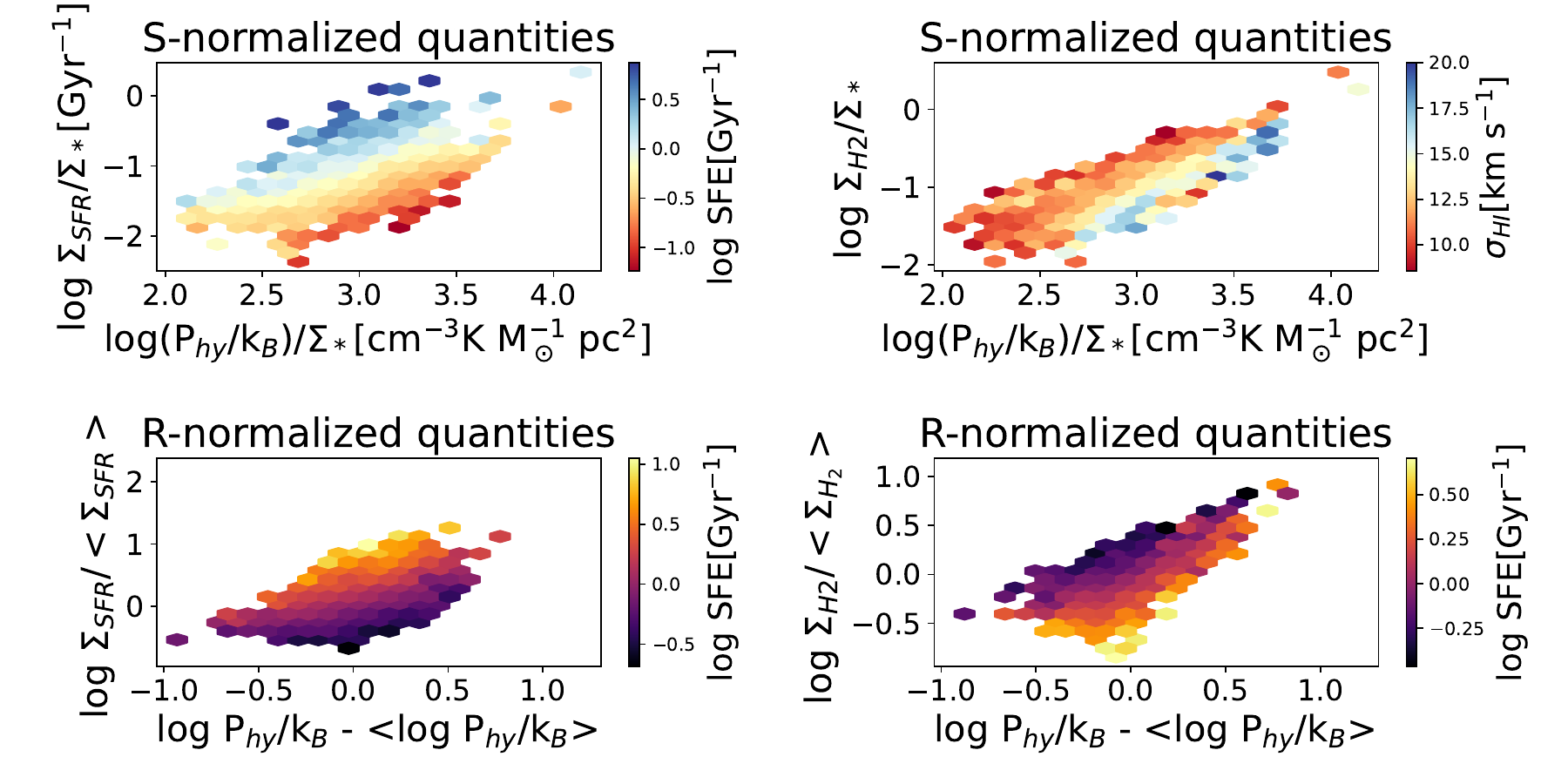}
\caption{Top panels: log $\Sigma_{SFR}$--log P$_{hy}$ and  log $\Sigma_{H_2}$--log P$_{hy}$ relations using S-normalized quantities. These are color-coded with the star formation efficiency (log SFE) or with the gas dispersion ($\sigma_{HI}$). Bottom panels:\ Same R-normalized relations, are all color coded with the SFE.}
\label{radnorm}
\end{figure*}

Normalizing physical quantities to the local  stellar surface density is also a way of removing radial variations because the stellar surface density scales with R, displaying marginal azimuthal variations. We  refer to  these as S-normalized quantities. In Figure~\ref{radnorm}, we show   R- and S-normalized correlations between the surface density of star formation rate  and pressure (left) and molecular mass surface density and pressure (right), using the effective SFE and HI velocity dispersion, $\sigma_{HI}$, as colors.  On the left, we can see that the  spread in normalized SFR at a given normalized pressure is almost entirely the result of a spread in  effective SFE.
This follows if pressure is the main driver of molecule formation, so that more or less star formation at a given pressure implies more or less star formation at a given molecular mass, which is essentially the definition of SFE. Similarly for the top-right panel, a higher HI turbulent speed for the same pressure corresponds to a lower molecular surface density, presumably because of the lower midplane density, which has a large impact on molecule formation. The lower-right panel resembles the lower left-panel in that $\Sigma_{H2}$ follows pressure with more or less star formation, corresponding inversely to the SFE. We note that the smallest vertical scatter occurs for the normalized molecular surface density at a given normalized pressure, suggesting that pressure-induced formation of molecules is the driving relationship --  even for azimuthal variations in star formation.

\subsection{Star formation drivers from RF analysis}

To   further  investigate what drives the SFR across the M33 disk, we used a RF analysis.   A RF regression can be used to test the relevance of several variables, some of which might be combinations of others \citep[e.g.,][for examining various combinations of $\Sigma_*$ and $\Sigma_{H_2}$]{2024MNRAS.52710201E}. Some caution should be used in comparing the relative importance of variables that are strongly correlated, especially if their relevance is not substantially different. 

\begin{figure} 
\includegraphics [width=9.cm]{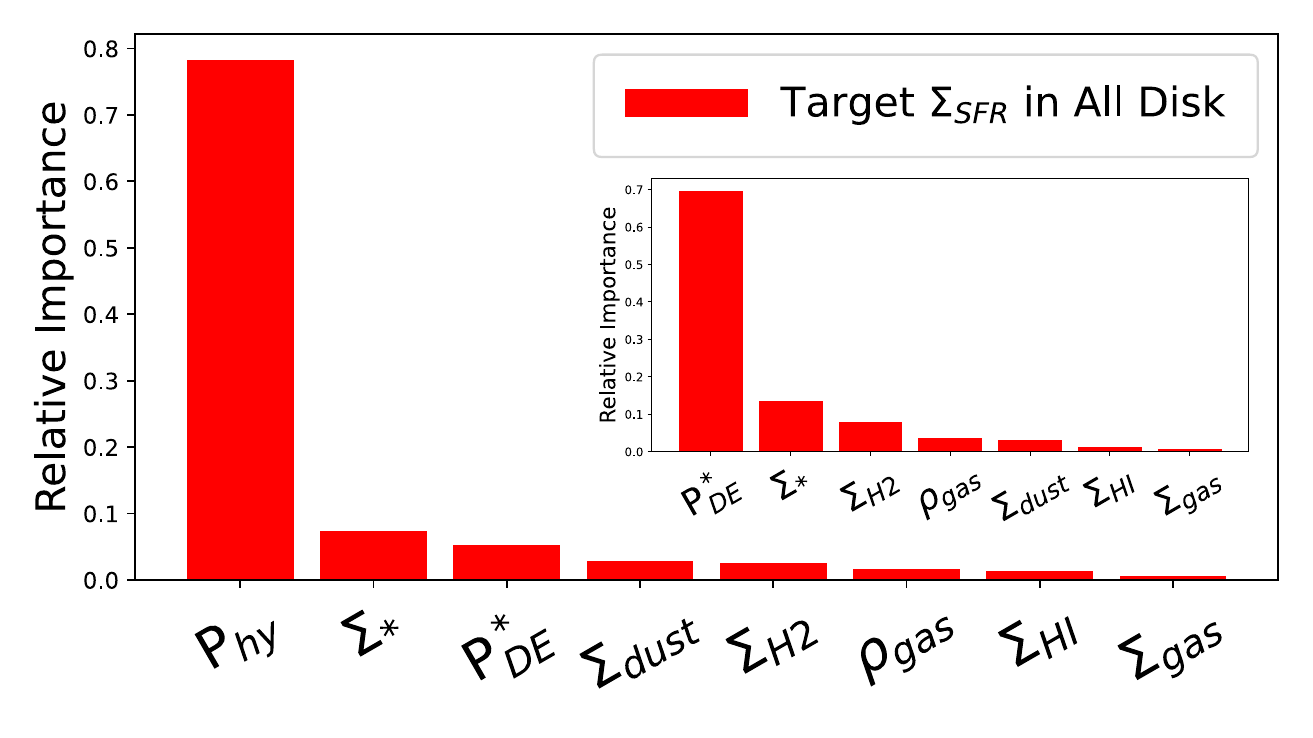}
\hspace{-0.5 cm}
\includegraphics [width=9.cm]{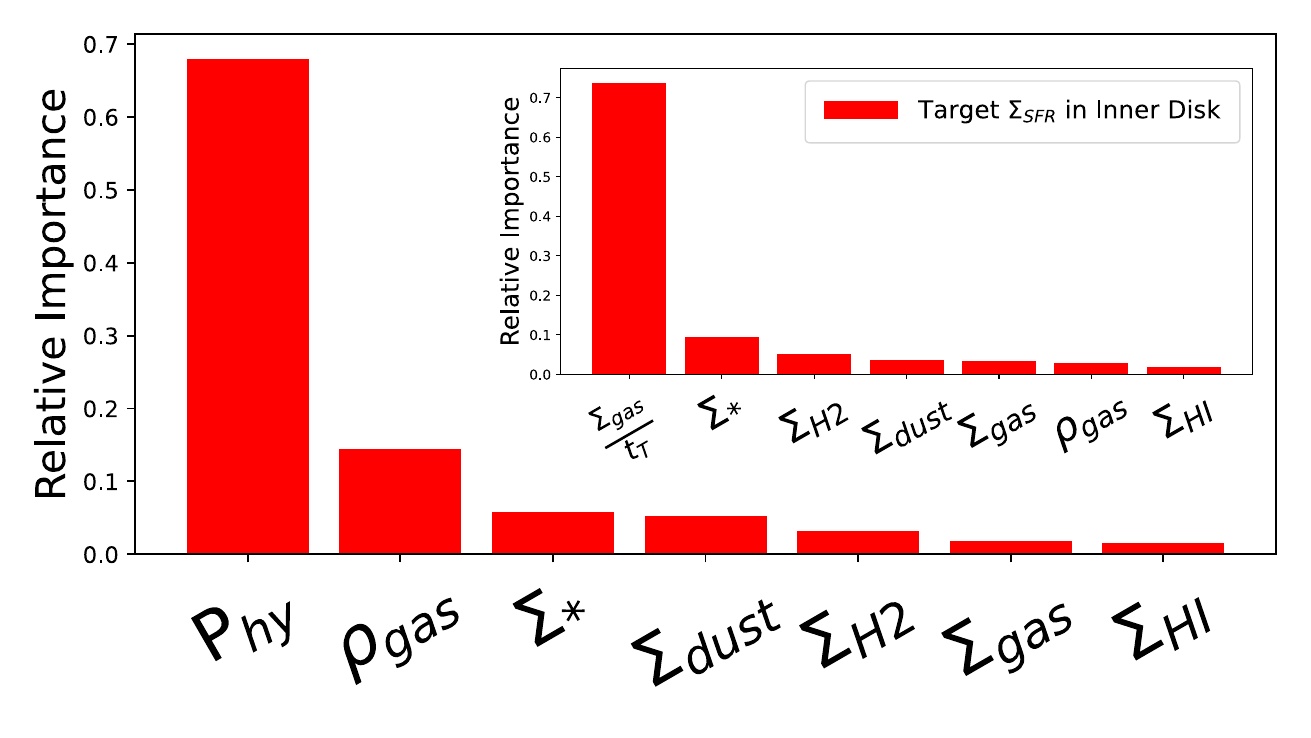}
\hspace{-0.5 cm}
\includegraphics [width=9.cm]{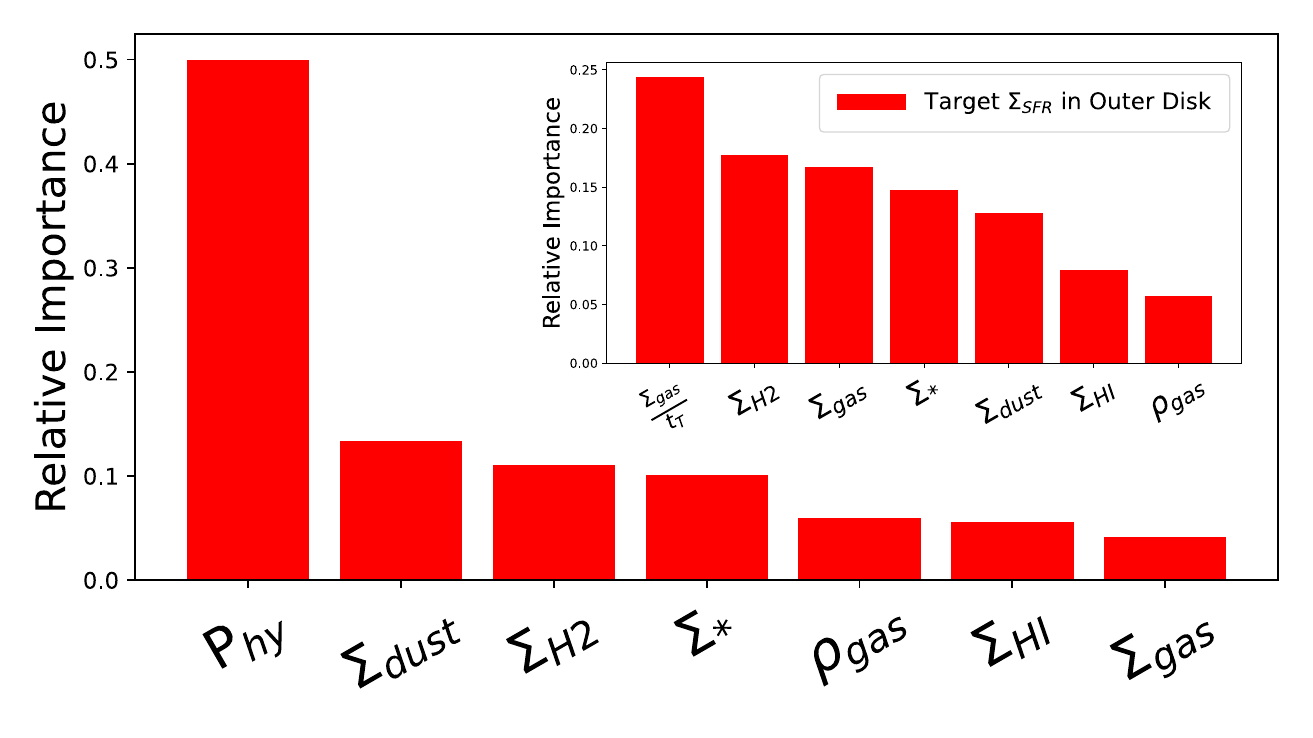}
\caption{High relative importance of P$_{hy}$ in predicting $\Sigma_{SFR}$, compared to other physical variables and determined from a RF regression analysis, is shown for the whole disk (top) and for the inner and outer disk separately (central and bottom panel, respectively).  We have considered also P$_{DE}^*$ for the whole disk analysis, which is the best $\Sigma_{SFR}$ predictor when P$_{hy}$ is not included, as shown by the small inset of the top panel. For the inner disk, the inset in the middle panel shown that for regions with Q<1 the Toomre condensation parameter, $\Sigma_{gas}/t_T$, strongly related to P$_{hy}$,  also has a high relative importance.  In the outer disk (bottom panel), P$_{hy}$ has the highest relative importance; however,  when we replace P$_{hy}$ with the Toomre condensation parameter, the small inset shows that this has a lower relevance than P$_{hy}$  for driving star formation.} 
\label{impo-dust}
\end{figure}

Using data for the whole star-forming disk we analyze the following variables $\Sigma_{HI}$, $\Sigma_{H_2}$, $\Sigma_{gas}$, $\Sigma_{*}$, $\Sigma_{dust}$, $\rho_{gas}$, P$_{DE}^*$, and  P$_{hy}$.  The relative importance of variables in predicting $\Sigma_{SFR}$ over the whole disk  or in the inner and outer disk separately is shown in Figure~\ref{impo-dust}. We find that hydrostatic pressure, P$_{hy}$, has the highest relative importance, reaching more than 75$\%$ in the whole star-forming disk as well as in the inner disk. All the panels confirm that P$_{hy}$ is a superior predictor of $\Sigma_{SFR}$ than the usual $\Sigma_{H_2}$ or $\Sigma_*$. We show importance of P$_{DE}^*$  in comparison to P$_{hy}$  in the top panel for all disk regression; here, P$_{hy}$ seems to be a superior predictor than P$_{DE}^*$, which is often used.    Because of the correlation between  P$_{hy}$ and P$_{DE}^*$ (especially in the inner disk), we also examined RF results keeping these two variables separately. Considering only P$_{DE}^*$,  we were able to recover the results of previous analyses, namely, that P$_{DE}^*$ has the highest importance to drive star formation.  This is shown in the inset of the upper panel of Figure~\ref{impo-dust}. Using the same set of variables but replacing  P$_{DE}^*$ with P$_{hy}$, the RF confirms that the relative importance of P$_{hy}$ is very high:\ above  80$\%$.
We  examined the inner and outer star-forming disk separately without considering further P$_{DE}^*$. In the central panel of Figure~\ref{impo-dust}, we see that the inner disk has marginal differences from the whole disk RF analysis. For  the outer disk, the bottom panel of  Figure~\ref {impo-dust}  shows  that  although P$_{hy}$  always has the highest level of importance, its relative importance is lower than in the inner disk: about 50$\%$. Here, other variables might contribute to drive the star formation activity, such as $\Sigma_{dust}$ and $\Sigma_{H_2}$. 

\begin{figure} 
\centering
\includegraphics [width=9.8 cm]{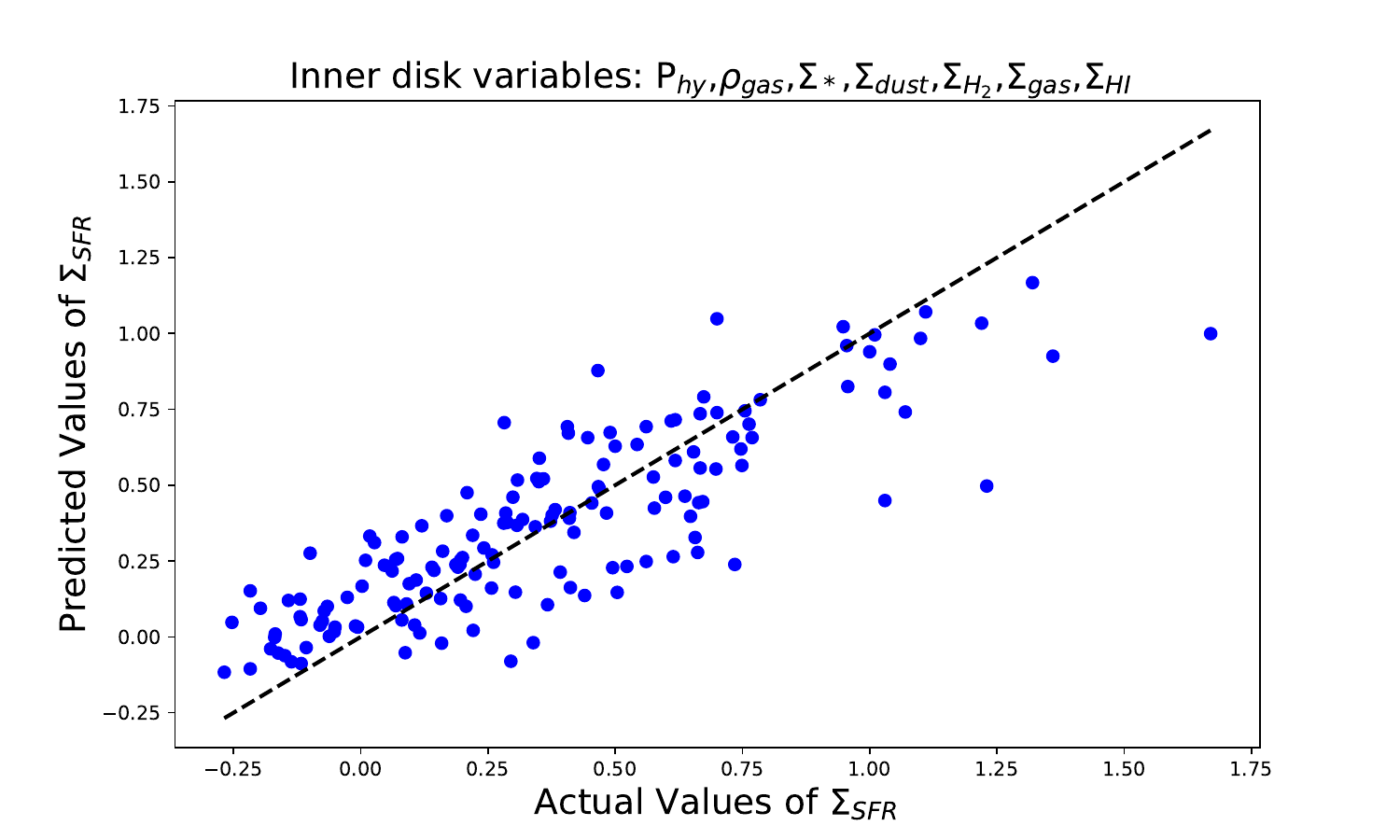}
\includegraphics [width=9.8 cm]{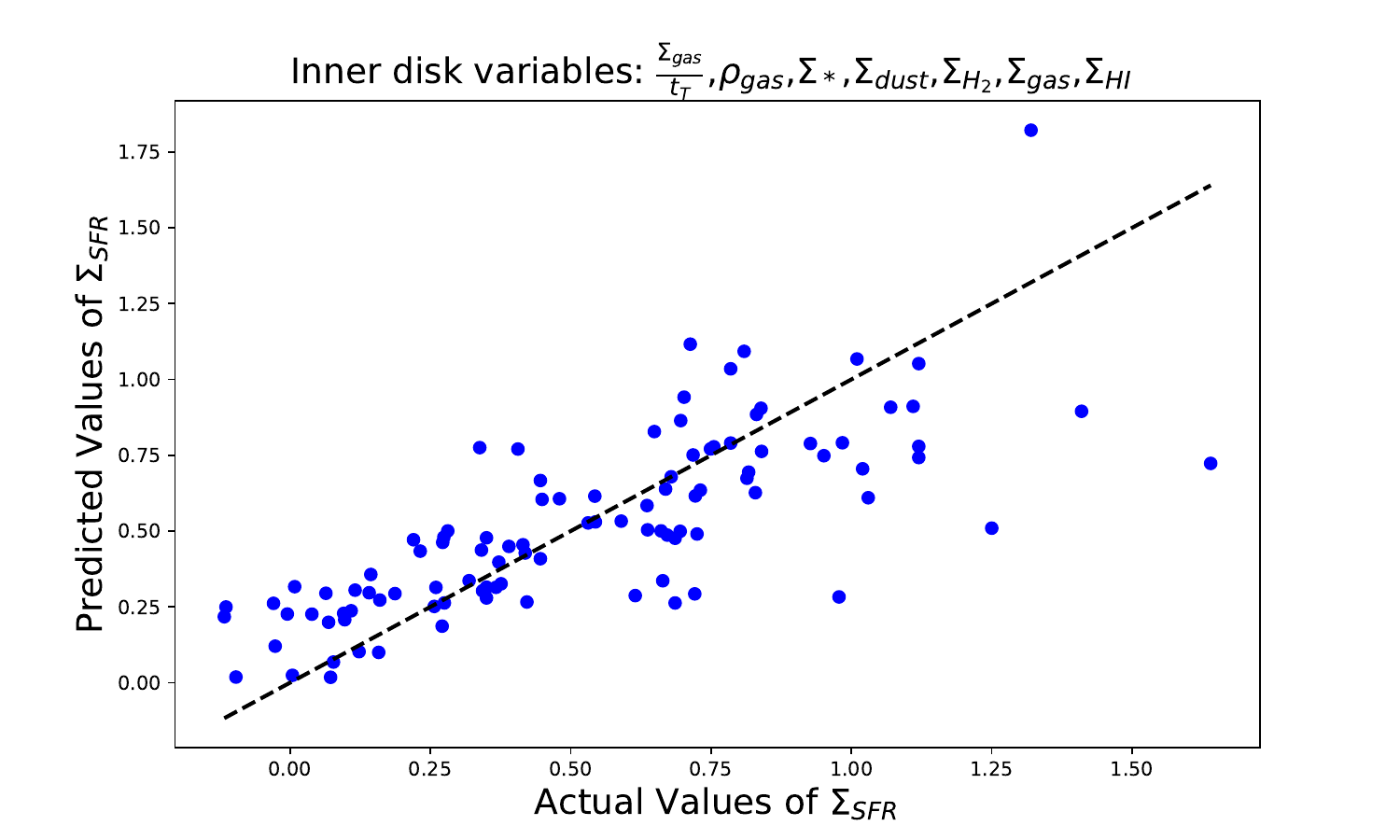}
\caption{  RF predicted values of $\Sigma_{SFR}$ versus its actual values  are shown for  the inner disk when the physical variables  listed above the panels are examined. In the top panel, all the inner disk data has been included, while  the bottom panel only shows  regions with Q<1. The relative importance of these variables are shown in the central panel of Figure~\ref{impo-dust}. The dashed line of unity slope has been drawn for reference. }
\label{predi}
\end{figure}

In the central and bottom panel insets of Figure~\ref{impo-dust}, we display the results of a RF regression selecting only regions for which Q $<1$ and replacing P$_{hy}$  with the Toomre condensation parameter, which is the ratio between the gas mass surface density and the characteristic Toomre timescale, $\Sigma_{gas}$/t$_T$  (see Appendix~\ref{apprad} for the definition of  Q, and t$_T$).  We consider the Toomre condensation as a test case for star formation drivers, possibly  related to the large scale disk motion. Simultaneously considering  the condensation parameter and P$_{hy}$, we find that they both have high relative importance, similar in strength. Given these two variables are strongly correlated across  the whole star-forming disk, as shown in Appendix~\ref{apprad}, it is hard to test their relative importance with the RF regression and we kept them separate. The  relative importance of P$_{hy}$ is the same as the Toomre condensation parameter when the same set of regions is examined  (Q<1, case not shown here).  It is likely that  azimuthal rotating disk perturbations regulate the growth of pressure enhancements where molecules form and star formation proceeds, especially in the inner disk which is dominated by the stellar potential. In Figure~\ref{predi} we show the good predictions of $\Sigma_{SFR}$   given by the RF for the inner disk.

In the outer disk,  P$_{hy}$ has the highest relative importance, shown in  the bottom panel of Figure~\ref{predi}, although this is lower than in the inner disk. For the outer disk, the predicted values of  $\Sigma_{SFR}$ decrease less than  the observed ones; namely, the combination of physical parameters tested predicts a star formation rate surface density which is lower than the observed one for high  $\Sigma_{SFR}$ and higher than the observed one for low  $\Sigma_{SFR}$. This implies a less accurate reproduction of the observed $\Sigma_{SFR}$  by the variables examined in this study.
The predictability of the distribution and the R$^2$ values of RF regression in the outer disk do not improve when considering the condensation parameter $\Sigma_{gas}$/t$_T$ in the RF analysis, as shown by the small inset  in the bottom panel of Figure~\ref{predi}. We chose not to investigate the Toomre parameter further in this work because our focus is on physical variables, rather than on the ratio of these with timescales.

\subsection{Zooming in, zooming out, and the PRFM theory}

\begin{figure} 
\hspace{-0.3 cm}
\includegraphics [width=9.3 cm]{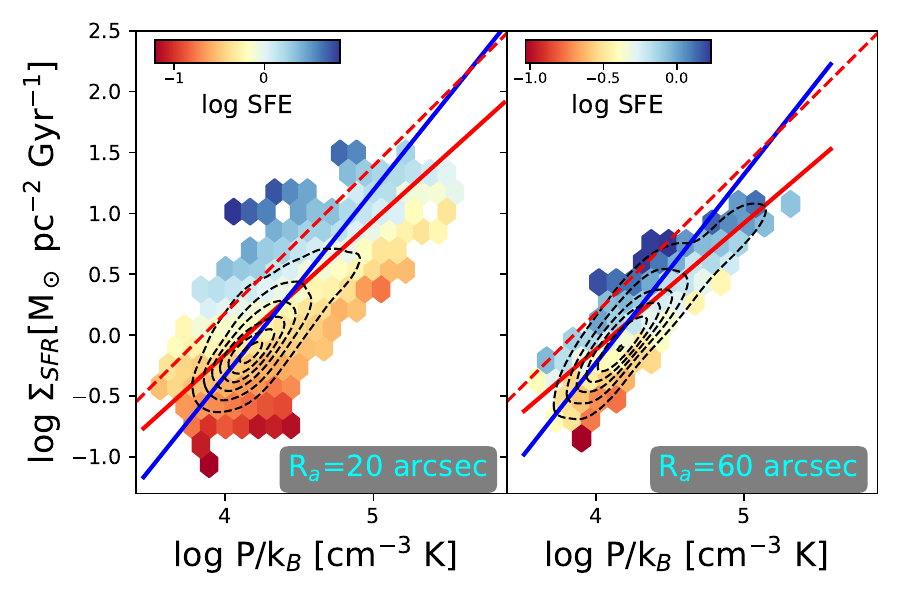}
\caption{log P$_{hy}$-- log $\Sigma_{SFR}$ relation is shown by varying the size of the sampled regions. Left panel: R$_{a}$ = 20~arcsec. Right panel: R$_{a}$= 60~arcsec. The dashed red line shows the relation predicted by the OK22 model. Color-coding is done with log SFE, which is the effective SFE per molecular mass surface density in Gyr$^{-1}$. The red and blue straight lines show the best fitted linear relations for ODR for the inner and outer disk, respectively. Dashed contours indicate the density of data in the plane. }
\label{aper}
\end{figure}

In this subsection, we investigate the effects of spatial resolution, that is, how the scaling relations change as we vary the size of the aperture. By zooming in or out, from apertures with 30~arcsec in radius to apertures with 20~arcsec or 60~arcsec in radius (i.e., from regions of 244~pc in size to regions of 163~pc and 488~pc in size), the slope variations of the scaling relations are small and compatible with the actual uncertainties. In  Figure~\ref{aper},  the log P$_{hy}$ -- log $\Sigma_{SFR}$ relation for these apertures is shown, color-coded with the SFE. The plotted linear relations refer to ODR results and both indicate a steepening of the slope going from the inner to the outer disk. Zooming out might be particularly relevant for the outer disk because the expected increase in the scale heights suggests that individual event of star formation might extend over larger areas. Correlations might, for example, improve by sampling at a spatial resolution that is comparable with the scale height. For this reason, we have  examined the scaling relations at spatial resolution as low as 1~kpc, comparable to the vertical scale height of the outermost star forming disk.  We find some slope variations but their larger uncertainties, due to the limited number of sampled regions at this resolution,  still keep them compatible with those quoted in Table~1. 

 By changing the aperture radius, the RF regression in both regions of the disk gives similar results to those discussed in the previous subsection, with P$_{hy}$ being the main driver but of minor significance in the outer disk. In the outer disk the  relative importance of P$_{hy}$ decreases by sampling M33 at lower resolution, while that of  $\Sigma_{dust}$ and $\Sigma_{H_2}$ increases. 

The dispersion around  the scaling relations increases as we increase the spatial resolution, as reported  in earlier works  for the molecular KS relation \citep[e.g.,][]{2010ApJ...722L.127O,2010A&A...510A..64V}. This is because the dispersion has a major contribution from  the star formation timescale and cycle (see Section 4.2). The dependence of scaling slopes on the spatial scale has been also discussed, for the molecular KS law only, in some studies dedicated to M33  \citet[e.g.][]{2010A&A...510A..64V,2018MNRAS.479..297W}. However, a detailed comparison is limited by the use of different star formation tracers than that adopted in this study, by the use of gas maps of lower sensitivity, or by the missing uncertainties on the scaling relations. The steeper molecular KS slopes we found for the outer disk also underline the relevance of using accurate uncertainties for scaling laws which include faint regions. We underline that we tested our results with two different statistical fitting methods and used also RF analysis for the relevance of physical variables to star formation process across the M33 disk.

We conclude this section by comparing the total, inner, and outer slope of the log P$_{hy}$-- log $\Sigma_{SFR}$ in M33 at 1~kpc resolution with the  midplane pressure and the star formation rate surface density relation predicted by the pressure-regulated feedback-modulated model (referred to as the OK22 model), which employs a similar spatial resolution. We recall the OK22 model predicts the following relation:

\begin{equation}
{\hbox{log}} \Sigma_{SFR} = 1.21\ {\hbox{log}} \Bigl({P\over k_B}\Bigr)  -7.66 
.\end{equation}

The all disk linear slope of log P$_{hy}$-log $\Sigma_{SFR}$ relation agrees with the OK22 model prediction being 1.24$\pm$0.07.  
However, at spatial resolution higher than 1~kpc  the inner slope is generally shallower than what OK22 model predicts, while the outer disk slope is steeper, suggesting some  deviation from the model. At a resolution of 1 kpc (which is more appropriate for this comparison), the larger slope uncertainties do not allow us to draw a definitive conclusion for the inner and outer disk separately.  The $\Sigma_{SFR}$ is, however, lower than that predicted by the model by about 0.3 dex (as shown in Figure~\ref{aper}).  This offset might be related to the feedback efficiency, which might increase in low-surface-density regions (OK22) as in the M33 disk. The alternative would be an effective lower pressure, which, however, would require a decrease in the gas dispersion to 3-4~km~s$^{-1}$, values below those expected from thermal broadening of the warm HI phase.   The pressure and star formation densities estimated in M33 are similar to those estimated by \citet{2018ApJ...855....7H} in dwarf galaxies.

\section{Summary and conclusions}

In this work, we  investigated possible drivers of star formation across the M33 disk by relating physical quantities in resolved regions of this galaxy to the star formation rate surface density traced by a combination of H$\alpha$ and 24$\mu$m emission. We  computed how the midplane gas pressure, in equilibrium with the weight of the disk baryonic components and dark matter, varies radially across the disk. In Appendix~\ref{appsh}, we tested various analytic approximations used to estimate mean gas densities and scale heights as a function of galactocentric distance. We  then used the resulting  best analytical approximations to compute the local pressure, scale height, and midplane density. With M33 being the closest nearby blue spiral galaxy, with no signs of interactions and no bulge \citep{2007ApJ...669..315C, 2024A&A...685A..38C}, we have been able to test star formation drivers  by  sampling its undisturbed disk at spatial scales from 1~kpc  down to about   160~pc. We can summarize our main results as follows:

\begin{itemize}
\item[\textbullet]
{The radial trends of ISM related quantities and the dynamical analysis of M33  clearly show two regimes: an inner disk  where the pressure is dominated by  stellar gravity and an outer disk, where the gas and dark matter  mass exceed the stellar mass and dominate the local pressure. The transition between the inner and outer disk is at R$\simeq 4$~kpc, which is close to  corotation where the main spiral arms end.
}
\item[\textbullet]
{The local vertical weight, which sets the local equilibrium pressure, P$_{hy}$, appears to be a primary driver of molecule formation  throughout the whole star-forming disk of M33. This hydrostatic pressure has  a tighter relation with $\Sigma_{H_2}$  than it has with $\Sigma_{SFR}$ (Fig. \ref{pan9}) because of additional variations of $\Sigma_{SFR}$  with  the effective SFE. The relations that regulates star formation throughout  the M33 disk can be expressed as
\begin{equation}
{\hbox {log}}\  \Sigma_{H_2} =0.8\  {\hbox {log}}\ \Bigl({\frac{P_{hy}}{k_B}}\Bigr)-3.0
,\end{equation}

\begin{equation}
{\hbox {log}}\  \Sigma_{SFR} = 1.0\  {\hbox {log}}\ \Bigl({\frac{P_{hy}}{k_B}}\Bigr)-4.1 \qquad {\hbox {R}}\le 4\ {\hbox {kpc}}
\label{sfr_p}
.\end{equation}
\begin{equation}
{\hbox {log}}\  \Sigma_{SFR} = 1.4\  {\hbox {log}}\ \Bigl({\frac{P_{hy}}{k_B}}\Bigr)-6.0 \qquad {\hbox {R}}> 4\ {\hbox {kpc}}
\label{sfr_p}
.\end{equation}

The slope of the $\Sigma_{SFR}$ versus $P_{hy}$ relationship is shallower in the inner disk than in the outer disk,  in contrast to  the $\Sigma_{H_2}$ versus $P_{hy}$ scaling, which is more uniform across the disk. The inner disk scaling of $\Sigma_{SFR}$ with P$_{hy}$ is close to linear. The steeper outer disk relation may be related to diffuse molecular gas. Here, small pressure changes can make a larger difference in converting the dominant atomic gas into a dense molecular medium for the formation of stars. 

The relation between $\Sigma_{H_2}$ and $\Sigma_{SFR}$ (KS law) is superlinear and has a larger dispersion. By examining regions from 163~pc to about 1~kpc in size, we did not find a clear physical scale dependence of the scaling laws examined. 

}

\item[\textbullet]
{We propose that variations in SFE have a large contribution from cloud evolution, where the number of new stars and their associated luminosities and ionizations increase over time, without significantly depleting the molecular mass. This follows from the use of these luminosities and ionizations as a SFR\ indicator, when in fact they only indicate the mass of stars formed up to the present time. Such an interpretation of SFE is useful for highly resolved regions as those we observe here.  In addition to the star formation timescale, the P$_{hy}$--$\Sigma_{SFR}$ relation shows also an additional  dependence on the gas dispersion that gives  a three-parameter correlation.
}
            
\item[\textbullet]
{The RF regression underlines that P$_{hy}$ is more relevant than the gas density and other variables in driving the star formation rate surface density throughout the whole disk, as also underlined  by correlation analyses, ODR, and BAY fits  shown in Table~1. The hydrostatic pressure is a superior predictor of $\Sigma_{SFR}$ than the usual $\Sigma_{H_2}$ or $\Sigma_*$. The relative importance of P$_{hy}$   and the accuracy of the predicted values are higher in the inner disk than in the outer disk. 
 }

 \item[\textbullet]
{By neglecting radial variations (using normalized quantities), we confirm that even local fluctuations have power-law  $\Sigma_{SFR}$--P$_{hy}$ and near-linear P$_{hy}$ -$\Sigma_{H_2}$ relationships, with fluctuations primarily from a varying effective SFE (or star formation timescale) and gas dispersion. 
}

\item[\textbullet]
{ The use of its full analytic expression for the midplane pressure, P$_{hy}$, makes it  a better predictor of the star formation rate surface density across the disk of M33 than more approximate expressions. 
 }
 
\end{itemize}

\begin{figure} 
 \includegraphics [width=9. cm]{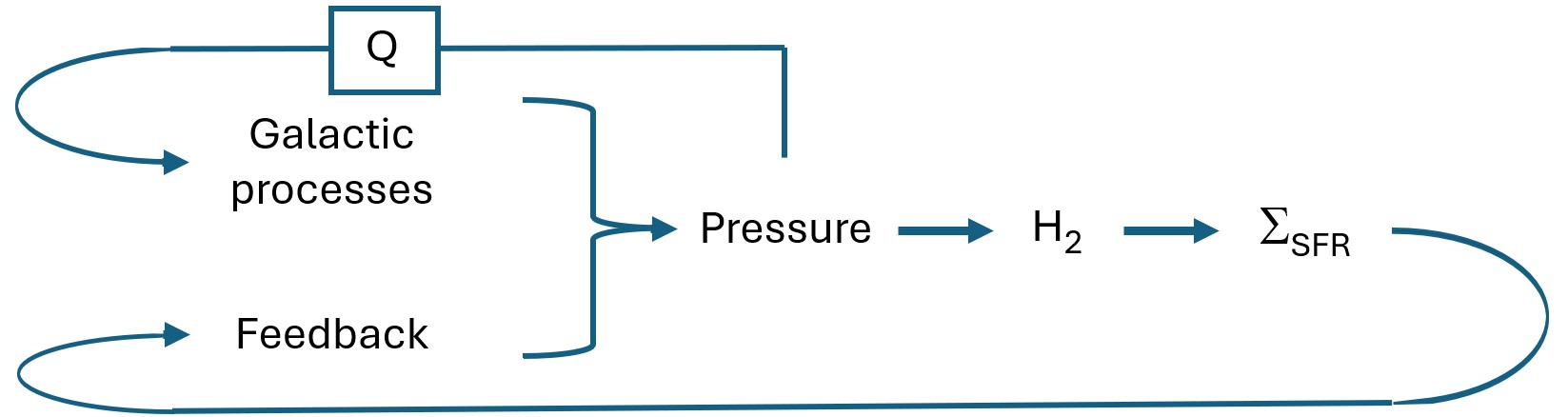}
\caption{A schematic view of the star formation cycle in M33}
\label{sfr_driver}
\end{figure}

These correlations for M33 confirm and improve upon many previous studies that suggest the primary driver for star formation is interstellar pressure \citep[e.g.,][]{2002ApJ...569..157W} and that the role of pressure is as a regulator for molecular gas \citep[e.g.,][]{1993ApJ...411..170E,2004ApJ...612L..29B}. We show here that the midplane pressure is determined primarily by the product of the gas surface density and the total mass of gas, stars, and dark matter inside the gas layer. This pressure presumably determines the density of diffuse clouds near the midplane and for higher values of pressure, causing more of these clouds to self-shield and form molecules, and to form these molecules faster with the higher densities. The formation of molecules, particularly CO, then leads to a temperature drop by a factor of $\sim10$; namely,  from the equilibrium temperature of the cool neutral medium to the equilibrium temperature of a molecular cloud. That drop, in turn, lowers the gravitational Jeans mass at this pressure by a factor of $\sim100$, which allows the cloud fragments and filaments, previously formed by turbulent compression, to collapse gravitationally and form stars. This sequence of events is represented in Figure \ref{sfr_driver} by the arrows from pressure to $H_2$ to $\Sigma_{SFR}$ and has become a standard model.   

The precursor to pressure in Figure \ref{sfr_driver} is the motion generated by star formation feedback on small scales and galactic processes such as spiral waves and gravitational instabilities on large scales. These combined motions control the thickness of the gas layer and therefore the total mass inside this layer, as faster motions lead to larger thicknesses that include more stars and dark matter in the layer.  This disk-breathing mode is often discussed as a regulator for instabilities and star formation \cite[e.g.,][]{1965MNRAS.130..125G,2010ApJ...721..975O}. Its combination with the tight relationships found here between pressure and H$_2$ and between the pressure and $\Sigma_{SFR}$ support the standard model. 

There is an additional control cycle between gravitational processes and pressure that involves the multi-component Toomre Q parameter, shown to be approximately constant in galaxies by \cite{2017MNRAS.469..286R}.  This cycle, also depicted in Figure \ref{sfr_driver}, regulates the velocity dispersion through gravitational instabilities \citep[e.g.,][]{2002ApJ...577..197W} and introduces a dependence of the SFR per unit gas on galactic dynamical rates, such as the epicyclic, Toomre, and orbital rates \citep[e.g.,][]{2024ApJ...966..233E}.
The presence of two control cycles related and connected to pressure makes the origin of the turbulence in any one region difficult to determine, especially when motions from these cycles are combined with additional motions internal to the clouds from self-gravity \citep[e.g.,][]{2024MNRAS.534.1043B} and cloud collisions \citep[e.g.,][]{2021PASJ...73S...1F}.

\begin{acknowledgements}
We acknowledge financial support from INAF-Mini Grant RF-2023-SHAPES. We would like to acknowledge Michele Ginolfi for his help in the random forest analysis, and the anonymous referee for his/her constructive report which has improved the clarity and science content of the original manuscript. 

\end{acknowledgements}

\begin{appendix}

\section{Uncertainties in photometric measurements}
\label{apperr}

The uncertainties for the physical quantities involved in our study are the following:
\begin{itemize}
    \item $\Sigma_{HI}$: we use the average rms in the 21-cm spectra (2.1 K per channel) for a typical linewidth of 30~km~s$^{-1}$  and a channel width of 1.25~km~s$^{-1}$. We then scale the noise according to the aperture-to-beam area ratio;
    \item $\Sigma_{H_2}$: we use the average rms of the integrated CO spectra (0.22~K~km~s$^{-1}$ \citep{2014A&A...567A.118D} in main beam temperature units). We  scale this noise according to the aperture-to-beam area ratio. We consider only regions where the integrated signal to  spectral noise ratio is $>1.5$ in the CO map. To this noise we add 20$\%$ uncertainties due to local variation of the CO-to-H$_2$ conversion and of the CO 2-1/1-0 line ratio;
    \item $\Sigma_*$: the optical stellar map has uncertainties of the order of 30$\% $ \citep{2014A&A...572A..23C}  at a resolution of 25~arcsec.  We assume similar uncertainties for the NIR stellar map and scale them with aperture-to-beam area ratio.. To these uncertainties we add in quadrature the dispersion around the average value of $\Sigma_*$ given by the optical and NIR stellar maps in each aperture;    
    \item $\sigma_{HI}$ and $\sigma_{H_2}$: we consider twice the channel width (i.e. 2.5~km~s$^{-1}$) as uncertainty for $\sigma_{HI}$. For $\sigma_*$ we  consider an uncertainty of 2~km~s$^{-1}$ as given by the dispersion of the measures for the old stellar population \citep{2022AJ....163..166Q};
      \item P$_{hy}$: we propagate the uncertainties of all physical quantities involved in the computation of P$_{hy}$ considering 10$\%$ additional uncertainties due to the dark matter surface density;
    \item $\Sigma_{SFR}$: calibration uncertainties of the H$\alpha$+24$\mu$m tracer  have been  estimated of the order  of  30$\%$ \citep{2007ApJ...666..870C}. Because we are sampling also fainter regions than those examined by \citet{2007ApJ...666..870C},  additional uncertainties  as high as 50$\%$ were added in quadrature to  previous estimates when the total H$\alpha$ luminosity in an aperture is smaller than 3$\times$10$^{37}$~erg~s$^{-1}$.  In this regime, in fact, the stochastic  birth of massive stars  \citep{2009A&A...495..479C} implies large fluctuations in the H$\alpha$ to stellar mass ratio \citep{2010A&A...521A..41G} giving a total uncertainty on the SFR density close to 60$\%$.
    
\end{itemize}

Uncertainties on the dust mass surface density have not been used in this paper but have been estimated to be of the order of 10$\%$.

\section{Scale heights and the local weight}
\label{appsh}

In this appendix, we compare the values retrieved from commonly adopted analytic expressions of the local vertical weight and scale heights in a galactic disk with  computational values using M33 as a test case. We consider the possibility that dark matter contributes to the local weight, and discuss possible estimates of the dark matter density for a generic galaxy. 

We consider the scale heights and weights  experienced by gas and stars in the midplane for a disk of gas, stars, and dark matter with total surface densities perpendicular to the galactic plane of $\Sigma_g$, $\Sigma_s$ and $\Sigma_{dm}$ respectively. These are functions of galactocentric radius, R, the distance of the midplane from the galaxy center. We assume a plane-parallel distribution of stars, gas, and dark matter and use M33 data as a test case.  A more complete theory would include vertical forces from a bulge or other mass concentration far from the position at $R$ \citep[e.g.,][]{2019MNRAS.484...81P,2019A&A...622A..64B,2022MNRAS.514.3329M}, but M33 has a negligible bulge so a local, plane-parallel approximation is sufficient.

The vertical distributions of gas and stars are determined numerically from four observables: $\Sigma_g$, $\Sigma_s$, $c_g$, $c_s$, and the dark matter density obtained from the rotation curve, $\rho_{dm}$. We assume the gaseous and stellar velocity dispersions, $c_g$ and $c_s$, are constant with height. Following \citet{2019ApJ...882....5W}, we include the effect of magnetic pressure and cosmic rays by multiplying the gas velocity dispersion by $\sqrt{1.3}$. In M33 the CO and HI dispersions are similar and the galaxy is dominated by atomic gas; we have therefore used $c_g=\sqrt{1.3} \sigma_{HI}$. The stellar velocity dispersion for M33 has been considered constant throughout the disk and equal to c$_s$=18~km~s$^{-1}$ \citep{2022AJ....163..166Q}. Then, for each component, the gradient of the pressure at some height $z$, is equal to the gravitational force density from all of the mass components between $-z$ and $z$. Dividing this equality by density to obtain a derivative of the log, we obtain
\begin{equation}
c_i^2{{d\log\rho_i}\over{dz}}=-2\pi G\int_{-z}^{z}\left(\rho_g+\rho_s+\rho_{dm}\right)dz
,\end{equation}
for $i$ representing gas ``g'' or stars ``s.''  The gravitational constant is $G$. The equations were solved by assuming central densities, integrating to very large heights, comparing the resultant surface densities to the observed values, and then correcting the central densities for a second integration over height. This procedure continued until the calculated and observed surface densities were equal. Because $\rho_{dm}$ does not vary significantly with height inside the gas and star layers, we assumed it to be constant. The midplane gas pressure resulting from this integration is designated $P_{true}=\rho_gc_g^2$, where we refer below to these computational values as true values. The true scale heights, h$_{i}^{true}$, are computed as the ratio of the mass surface density to twice the central density, $h_i=\Sigma_i/(2\rho_i[z=0])$.

For simple analytical approximations to the midplane pressure and density, we assume that the gas velocity dispersion is smaller than the stellar velocity dispersion, and therefore the gas scale height is smaller than the stellar one, h$_g <$ h$_s$, and the gas and stellar densities perpendicular to the disk have a distribution similar to that of an isothermal layer in vertical equilibrium, as in the ``true'' solution. A one-component solution  in this case will follow 1/cosh$^2(z/h)$.  Near the midplane, this is close to a Gaussian function  with a dispersion of $h/\sqrt(2)$,  and far from the midplane, it is close to an exponential function with a scale height of $2h$.  

In the presence of multiple components we approximate their weights, as 

\begin{equation}
W_g={\pi G \over 2} \Sigma_g \Sigma_t^g = {\pi G \over 2} \Sigma_g \lbrack \Sigma_g + \Sigma_s^g + \Sigma_{dm}^g \rbrack 
,\end{equation}

\begin{equation}
W_s={\pi G \over 2} \Sigma_s \Sigma_t^s = { \pi G \over 2} \Sigma_s \lbrack \Sigma_g + \Sigma_s + \Sigma_{dm}^s \rbrack   
,\end{equation}

\noindent
where $\Sigma_s^g$ and $\Sigma_{dm}^g$ are the stellar and dark matter surface density, respectively, within the gas layer, and  $\Sigma_{dm}^s$ is  the  dark matter surface density within the stellar layer. 
In equilibrium the weights are balanced by their local pressures,    P=$\rho_i c_i^2$. We call P$_{hy}$ the midplane pressure experienced by the gas. Using the equilibrium condition in the midplane, where the gas and stellar densities are $\rho_g$ and $\rho_s$  respectively, we define the stellar scale heights  h${_g,s} \equiv \Sigma_{g,s}/(2\rho_{g,s})$ and we can write:

\begin{equation}
W_g=\rho_g c_g^2 \equiv P_{hy} \qquad W_s = \rho_s c_s^2 
,\end{equation}

\begin{equation}
h_g={\Sigma_g \over 2 \rho_g}={\Sigma_g c^2_g \over 2 W_g} \qquad h_s={\Sigma_s \over 2 \rho_s} = {\Sigma_s c^2_s \over 2 W_s}
.\end{equation}

\noindent
These scale heights are the equivalent to the constant h in the argument of the hyperbolic cosine for an isothermal solution. We recall  that this is half of the exponential scale height at large distances from midplane and  lower by a factor 1.085 than the height where the density drop to 1/e of its midplane value.
To find the scale height values we need to estimate $\Sigma_s^g$, $\Sigma_{dm}^g$, $\Sigma_{dm}^s$. As a zero order approximation we neglect dark matter, and write $\Sigma_s^g$=$\Sigma_s$ (c$_g$/c$_s$). In this case we recover the usual expressions for scale heights and midplane pressure as given by \citet{1989ApJ...338..178E} expressed by

\begin{equation}
h_g^0 = {c_g^2} \Bigl\lbrack {\pi G } \bigl(\Sigma_g+{c_g\over c_s}\Sigma_s\bigr) \Bigr\rbrack^{-1}  \qquad  h_s^0 = { c_s^2} \Bigl\lbrack {\pi G} \bigl(\Sigma_s+\Sigma_g\bigr) \Bigr\rbrack^{-1}  
,\end{equation}

\begin{equation}
P_{hy}^0 = {\pi G \over 2} \Sigma_g \Bigl\lbrack \Sigma_g + {c_g\over c_s} \Sigma_s  \Bigr\rbrack 
.\end{equation}

To improve our estimate, we consider  in the above equations the additional contribution of the dark matter. The radial profile of the dark matter density in M33 has been inferred through the dynamical analysis of the rotation curve. Through this analysis \citet{2014A&A...572A..23C} found  that a Navarro Frank and White (NFW) halo model \citep{1997ApJ...490..493N} provides the best fit to the M33 data for a concentration parameter  C=9.5 and a halo mass of 4$\times10^{11}$~M$_\odot$. With these parameters the scale radius of  the halo model is r$_s$=20.2 kpc, and variations of distances from the galaxy center are small as we move perpendicular to the galactic plane in the star-forming disk of M33. It is then a good approximation to consider dark matter with a constant volume density, equal to the midplane density, throughout the whole vertical extent of the baryons.  A first order approximation, which we shall call it $1d$, can then be derived by estimating the gravitational contribution of the stars to the gas gravity $\Sigma_s^g=\Sigma_s(c_g/c_s)$, we have  the following expressions:

\begin{equation}
h_g^{1d} = {c_g^2} \Bigl\lbrack {\pi G } \bigl(\Sigma_g+{c_g\over c_s}\Sigma_s + 2 \rho_{dm}^0\xi h_g^{1d} \bigr) \Bigr\rbrack^{-1}  
,\end{equation}
\begin{equation}
h_s^{1d} = { c_s^2} \Bigl\lbrack {\pi G } \bigl(\Sigma_s+\Sigma_g + 2 \rho_{dm}^0\xi h_s^{1d} \bigr ) \Bigr\rbrack^{-1}  
,\end{equation}

\begin{equation}
P_{hy}^{1d} = {\pi G \over 2} \Sigma_g \Bigl\lbrack \Sigma_g + {c_g\over c_s} \Sigma_s  + 2 \rho_{dm}^0\xi h^{1d}_g \Bigr\rbrack 
,\end{equation}
where the factor $\xi$ accounts for the deviations of the vertical distribution of dark matter from an isothermal distribution. For an isothermal layer with central density $\rho_0$, the mass surface density between -z and z is  $\Sigma(z)=2 h \rho_0$ tanh(z/h) and  integrating  between -h and h we recover 76$\%$ of the total mass. 
For this reason we estimate $\xi$=1/0.76=1.32 and write the dark matter gravitational contribution as $\Sigma_{dm}= 2\rho_{dm} h_{g,s}$ with $\rho_{dm}=\rho_{dm}^0$ /0.76, where h$_{g,s}$=h$_g$ or h$_{g,s}$=$h_s$ according to whether we consider the equilibrium of gas or stars. 
The two scale heights, solutions of second order algebraic equations, and the midplane pressure can then be written as:

\begin{equation}
h_g^{1d} ={2 c_g^2 /\pi G \over (\Sigma_g +{c_g\over c_s} \Sigma_s)+ \sqrt{(\Sigma_g +{c_g\over c_s} \Sigma_s)^2+ {8 c_g^2  \rho_{dm} \over \pi G}  }}
,\end{equation}

\begin{equation}
h_s^{1d} ={2 c_s^2 /\pi G \over (\Sigma_g +\Sigma_s)+ \sqrt{(\Sigma_g +\Sigma_s)^2+ {8 c_s^2  \rho_{dm} \over \pi G}  }}
,\end{equation}

\begin{equation}
P_{hy}^{1d}={\pi G\Sigma^2_g \over 4}  \Biggl\lbrace 1+{ c_g\Sigma_s \over  c_s\Sigma_g} + \sqrt{\Bigl\lbrack 1+{ c_g\Sigma_s \over c_s\Sigma_g}\Bigr\rbrack^2+8 {c_g^2 \rho_{dm}\over \pi G \Sigma_g^2}  }\  \Biggr\rbrace
.\end{equation}

A second approximation which we consider  is based on using the scale heights themselves for estimating the  stellar gravity in the gas layer  $\Sigma_s^g$=$(h_g/h_s)\Sigma_s$. In this case the scale heights, labeled $2d$, are solutions of the following second order algebraic equations:

\begin{equation}
h_g^{2d} = {c_g^2} \Bigl\lbrack {\pi G} \bigl(\Sigma_g+{h^{2d}_g\over h^{2d}_s}\Sigma_s + 2\rho_{dm} h_g^{2d} \bigr) \Bigr\rbrack^{-1}  
,\end{equation}

\begin{equation}
h_s^{2d} = { c_s^2} \Bigl\lbrack {\pi G } \bigl(\Sigma_s+\Sigma_g +2 \rho_{dm} h_s^{2d}\bigr ) \Bigr\rbrack^{-1}  
,\end{equation}

\noindent
which give:

\begin{equation}
h_g^{2d} = {2 c_g^2/\pi G \over   \Biggl\lbrace \Sigma_g + \sqrt{\Sigma^2_g+{8 c_g^2\over \pi G }\Bigl\lbrack{ \rho_{dm} + \rho_s } \Bigr\rbrack }\  \Biggr\rbrace} \qquad  h_s^{2d} = h_s^{1d}
,\end{equation}

\begin{equation}
P_{hy}^{2d} = {\pi G\Sigma^2_g \over 4}  \Biggl\lbrace 1 + \sqrt{1+{8 c_g^2\over \pi G}\Bigl\lbrack{ \rho_{dm} + \rho_s \over  \Sigma_g^2} \Bigr\rbrack }\  \Biggr\rbrace = {\pi G \over 2} \Sigma_g^2 {\rho_g+\rho_s+\rho_{dm}\over \rho_g}
.\end{equation}

\noindent
In expressing the pressure we have used $\rho_s=\Sigma_s$/(2h$_s^{2d}$) and $\rho_g=\Sigma_g$/(2h$_g^{2d}$). 
To check the consistency of the assumption about the dark matter density, we  have also considered the effective dependence of the dark matter density from the height z=$\sqrt{r^2-R^2}$, where $R$ is the galactocentric radius in the midplane and $r$ is the distance from the galaxy center. We call this the iterative solution since it requires a numerical integration and iteration:

\begin{equation}
h_g^{iter} = {c_g^2} \Bigl\lbrack {\pi G} \Bigl(\Sigma_g+{h^{iter}_g\over h^{iter}_s}\Sigma_s + 2\int^{h_g^{iter}}_0{\rho_{dm}(z) dz} \Bigr) \Bigr\rbrack^{-1}  
,\end{equation}

\begin{equation}
h_s^{iter} = { c_s^2} \Bigl\lbrack {\pi G } \Bigl(\Sigma_s+\Sigma_g + 2\int^{h_s^{iter}}_0{\rho_{dm}(z) dz} \Bigr) \Bigr\rbrack^{-1}  
,\end{equation}

\begin{equation}
P_{hy}^{iter} = {\pi G \over 2} \Sigma_g \Bigl\lbrack \Sigma_g + {h^{iter}_g\over h^{iter}_s} \Sigma_s  + 2\int^{h_g^{iter}}_0{\rho_{dm}(z) dz}  \Bigr\rbrack 
.\end{equation}

\begin{figure*} 
\centering
\hspace{0.5 cm}
\hspace*{-0.5 cm}
\includegraphics [width=16. cm]{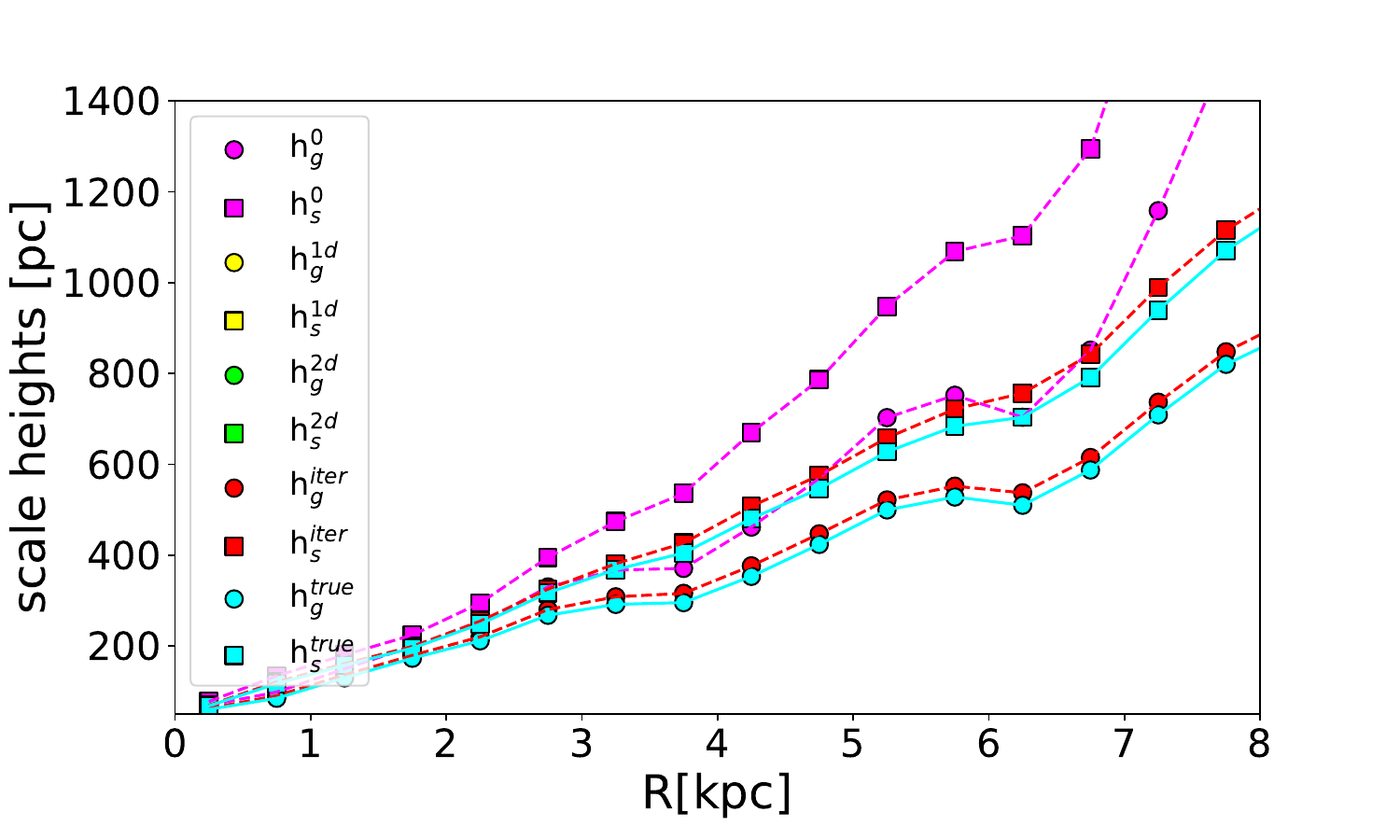}
\caption{Comparison of the   true  vertical scale heights of the matter distribution perpendicular to the disk of M33, as from numerical computation, with analytic expressions which can be used to estimate their values and which have been reported in this Appendix. Square symbols are for the stellar components, whose scale heights for $1d,2d$ and $iter$ approximation  overlap. Filled circles are for the gas components. }
\label{sh}
\end{figure*}

In Figure \ref{sh} we compare the values of h$^{true}$ to the  approximate values h$^0$, h$^{1d}$, $h^{2d}$, h$^{iter}$ for the gas and the stars, and we do similar comparison for the gas pressure.  In the pressure panel we added a few values relative to different estimates of the gas pressure which we explain in the following paragraph.

An analytical expression for the midplane gas pressure resulting from a vertical balance with gravitational forces has been suggested also by \citet{2004ApJ...612L..29B}, which we shall call  P$_{BR}$, and which is expressed as\begin{equation}
P_{BR} = {\pi G \over 2}  \Sigma^2_g +\Sigma_g c^0_g \sqrt{2G \rho_s}
,\end{equation}

\noindent
where c$^0_g$ is the gas velocity dispersion not  corrected for non thermal pressure. They suggest to write the central stellar density as $\rho_s= \pi G \Sigma_s^2/(2 c_s^2)$, as for a one component isothermal layer. In this case, however, we notice that the P$_{BR}$ expression is not equivalent to any of the above expressions for P$_{hy}$ with $\rho_{dm}=0$ and c$_g^0=c_g$. We compare it for example with $P_{hy}^0$ which we can rewrite as

\begin{equation}
P_{hy}^0 = {\pi G \over 2}  \Sigma^2_g + \Sigma_g c^0_g \sqrt{2 \rho_s G {\pi \Sigma_s\over 4(\Sigma_s+\Sigma_g)} }
.\end{equation}

\noindent
The second term of P$_{BR}$ is higher by a factor $\sqrt{4 (\Sigma_s+\Sigma_g)/(\pi \Sigma_s)} $ than P$_{hy}$ and even in the absence of gas the two expressions differ by a factor $\sqrt{4/\pi}$.
Many authors use P$_{BR}$  assuming  a constant vertical stellar  scale height  throughout the disk and we refer to this as P$_{BR}^0$ or equivalently as P$_{DE}^*$ (because it reproduces the  limiting expression of the dynamical equilibrium pressure P$_{DE}$ within 20$\%$ as defined later in this Appendix). The exponential scale height is usually related to the radial scale length of the stellar surface density \citep[e.g.][]{2008AJ....136.2782L,2010ApJ...721..975O,2020ApJ...892..148S,2021MNRAS.503.3643B,2024MNRAS.52710201E}  l$_s$, namely h$^{exp}_s$=l$_s$/7.3. Therefore  h$_s$= 2 l$_s$/7.3 and we have: 

\begin{equation}
P_{BR}^{0}\equiv P_{DE}^* = {\pi G \over 2}  \Sigma^2_g +\Sigma_g c^0_g \sqrt{2G \rho_s} \qquad {\hbox{with\ }} \rho_s\simeq{7.3 \Sigma_s\over 4 l_s }
.\end{equation}

\noindent
By adding dark matter and non thermal pressure to the expression of $P_{BR}$ and using a radially varying stellar scale height we have:

\begin{equation}
  P_{BR}^{d}  = {\pi G \over 2}  \Sigma^2_g +\Sigma_g c_g \sqrt{ 2G (\rho_s+\rho_{dm}) }
 .\end{equation}

\noindent
To estimate the stellar density BR04 suggest to use the one component isothermal solution and write $\rho_s=\pi G \Sigma_s^2/ (2 c_s^2)$. This  underestimates the stellar density in the presence of gas and dark matter and lacks of consistency with the overall treatment. However it allows the above equation to be be written as:

\begin{equation}
P_{BR}^{1d}  = {\pi G \over 2}  \Sigma^2_g \Biggl\lbrace 1+\sqrt{ {4 c_g^2 \Sigma_s^2 \over \pi  c_s^2 \Sigma_g^2} + 8{\rho_{dm}c_g^2 \over \pi^2 G \Sigma_g^2} }\  \Biggr\rbrace
.\end{equation}

\noindent
If we do a further refinement and write $\rho_s=\Sigma_s/(2h_s) = \pi G \Sigma_s \Sigma^s_t/(2 c_s^2) $ with $\Sigma^s_t=\Sigma_s+\Sigma_g+\Sigma_{dm}^s$ we can write

\begin{equation}
P_{BR}^{2d}  = {\pi G \over 2}  \Sigma^2_g \Biggl\lbrace 1+\sqrt{ {4 c_g^2 \Sigma_s \Sigma_t^s \over \pi  c_s^2 \Sigma_g^2} + 8{\rho_{dm}c_g^2 \over \pi^2 G \Sigma_g^2} }\  \Biggr\rbrace
,\end{equation}

\noindent
which  can be used only after evaluating the dark matter contribution to $\Sigma_t^s$ i.e. to the stellar scale height.
An analytic general expression for the midplane pressure has also been given by \citet{2011ApJ...743...25K}  and \citet{2022ApJ...936..137O} (OK22),  which they call P$_{DE}$, dynamical equilibrium pressure, expressed as

\begin{equation}
P_{DE}={\pi G\Sigma^2_g \over 4}  \Biggl\lbrace 1 + \sqrt{1+{32 c_g^2\over \pi^2 G}\Bigl\lbrack{ \rho_{dm} + \rho_s \over  \Sigma_g^2} \Bigr\rbrack }\  \Biggr\rbrace
.\end{equation}

\noindent
The above equation provides an analytic formula for the midplane pressure using the local vertical equilibrium of a disk  gas, stars and dark matter  (see OK22, their Eq.(6)), namely,  P$_{DE}$ has the same physical meaning as  P$_{hy}$ and we  make a comparison between the two and with the computational value.  As \citet{2011ApJ...743...25K}  and \citet{2022ApJ...936..137O} pointed out,  the two differ by less  than 20$\%$ and P$_{BR}^{d}$ (i.e., their Eq.(7)) is an approximation to P$_{DE}$ when the stellar component dominates over the gas component i.e. for $32 c_g^2 (\rho_{dm} + \rho_s)>> \pi^2 G  \Sigma_g^2$.

Unfortunately, the midplane stellar density in P$_{BR}$ or P$_{DE}$ is not an observable and an estimate of its value for spatially resolved analyses requires  an evaluation of  stellar scale heights across galaxy disks. We use the symbol P$_{DE}^0$ to indicate P$_{DE}$ computed for a radially constant scale height as in P$_{BR}^0$ with $\rho_s=\Sigma_s/(4 l_s/7.3)$:

\begin{equation}
P_{DE}^0={\pi G\Sigma^2_g \over 4}  \Biggl\lbrace 1 + \sqrt{1+{32 c_g^2\over \pi^2 G}\Bigl\lbrack{ \rho_{dm} \over  \Sigma_g^2} +  { 7.3 \Sigma_s \over 4 l_s \Sigma_g^2} \Bigr\rbrack}\  \Biggr\rbrace
.\end{equation}

\noindent
To avoid using a radially constant scale height for the stellar component in the P$_{DE}$, we express the stellar density as $\rho_s=\Sigma_s$/(2h$_s$) with h$_s$=h$_g(c_s/c_g)$.  Using eq(4) of OK22 with  $\zeta=1/\pi$, as the authors suggest, and solving a second order algebraic equation we derive the gas scale height h$_{g,DE}^{1d}$ and the corresponding estimate of the dynamical equilibrium pressure, which is expressed as

\begin{equation}
P_{DE}^{1d}={\pi G\Sigma^2_g \over 4}  \Biggl\lbrace 1+{4 c_g\Sigma_s \over \pi c_s\Sigma_g} + \sqrt{\Bigl\lbrack 1+{4 c_g\Sigma_s \over \pi c_s\Sigma_g}\Bigr\rbrack^2+32 {c_g^2 \rho_{dm}\over \pi^2 G \Sigma_g^2}  }\  \Biggr\rbrace
.\end{equation}

The P$_{DE}^{1d}$  can be easily compared to Eq.(6) of OK22 and see how P$_{DE}$ changes if one  replaces $\rho_s$ with the observable $\Sigma_s$ and c$_s$. Comparing  P$_{DE}^{1d}$  to P$_{hy}^{1d}$, one can see that  the term containing $\Sigma_g/\Sigma_s$ and $\rho_{dm}$ in P$_{DE}^{1d}$ have an extra 4/$\pi$ factor.
Using the more rigorous expression for the stellar density  in a multi-component layer, as for P$_{BR}^{1d}$, we have  $\rho_s=\pi G \Sigma_s \Sigma_t/(2 c_s^2)$. Using this expression for $\rho_s$ in Eq. (5) of OK22 we can estimate the gas scale height, h$_{g,DE}^{2d}$,  and the midplane pressure, expressed as

\begin{equation}
h^{2d}_{g,DE}= {2 c_g^2 \over  \pi G \Sigma_g \Biggl\lbrace 1 + \sqrt{ 1+ {32  c_g^2  \rho_{dm}\over  \pi^2 G\Sigma_g^2} +{16 c_g^2 \Sigma_s \Sigma_t^s \over \pi c_s^2 \Sigma_g^2} }\ \Biggr\rbrace }
,\end{equation}

\begin{equation}
P_{DE}^{2d}={\pi G\Sigma^2_g \over 4}  \Biggl\lbrace 1 + \sqrt{ 1+ 32 {c_g^2 \rho_{dm}\over \pi^2 G \Sigma_g^2 } + 16 {c_g^2 \Sigma_s \Sigma_t^s \over \pi c_s^2  \Sigma_g^2}  }\  \Biggr\rbrace
.\end{equation}

\noindent
To infer the dark matter contribution to the stellar scale height i.e. the $\Sigma_{dm}^s$ we use  $\Sigma_{dm}^s \simeq  2 \rho_{dm} h^{2d}_{g,DE}$ where in the h$_g^{2d}$ term we neglect  dark matter  in $\Sigma_t$. 
\begin{figure*} 
\hspace{-0.4 cm}
\includegraphics [width=10.3 cm]{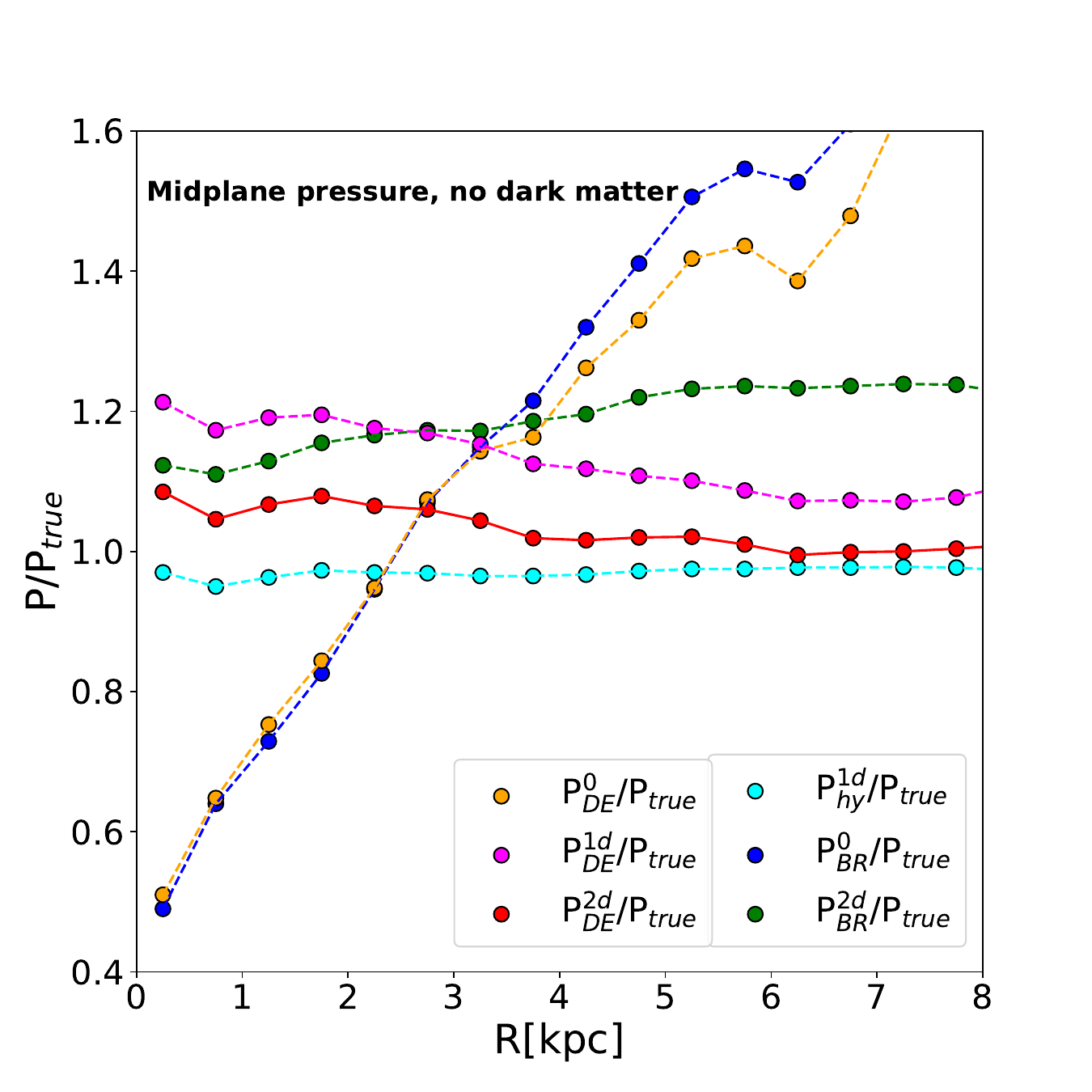}
\hspace*{-1. cm}
\includegraphics [width=10.3 cm]{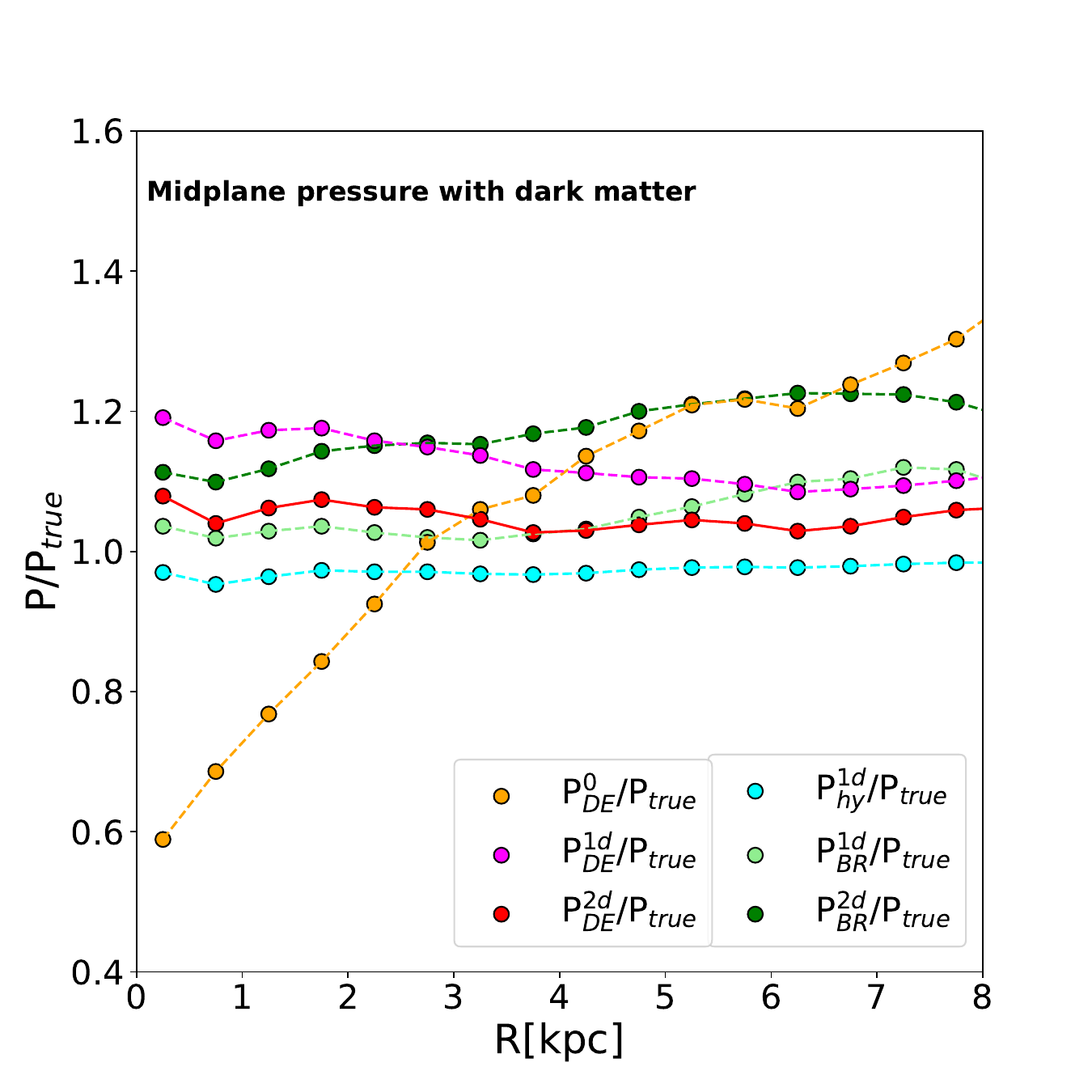}
\caption{Comparison of the   true weight of the matter distribution in the midplane of the M33 disk, P$_{true}$, as from numerical computation of equilibrium models, to analytic expressions  reported in this Appendix. In the left panel we plot the ratio of  analytic approximations of midplane pressure P to the true pressure P$_{true}$  for the case of no dark matter.  In the right panel we can see similar ratios  when dark matter has been considered in M33.  According to  color, P refers to P$_{hy}$, P$_{BR}$, and P$_{DE}$  expressions across the M33 disk (see text for details).}
\label{p+pdm}
\end{figure*}

Using for M33 l$_s=1.4$~kpc  in the Figure~\ref{p+pdm} we show the ratio of P$_{hy}$  P$_{DE}$ and P$_{BR}$ to P$_{true}$ with and without considering dark matter. We use P$_{hy}^{1d}$ as representative value of P$_{hy}$ with dark matter, given the very similar values of P$_{hy}^{1d}$, P$_{hy}^{2d}$ and P$_{hy}^{iter}$. For the case with no dark matter P$_{hy}^{0}$ = P$_{hy}^{1d}$. To make a comparison between P$_{BR}^{2d}$ and P$_{DE}^{2d}$ in the right panel of Fig.~\ref{p+pdm} we use  $\Sigma_{dm}^s \simeq 2 \rho_{dm} h^{2d}_{g,DE}$.

 It is clear that the use of radially constant scale height in P$^0_{DE}$ or P$^0_{BR}$ gives very discrepant results from the numerical solution of  hydrostatic equilibrium. 
This is independently of whether we consider or not dark matter in the disk.  Comparing the simplified expressions for a radially constant scale height in the case of no dark matter, P$_{BR}^0$ and P$_{DE}^0$, (see left panel of Figure~\ref{p+pdm})   we see that the Figure confirms they are rather similar  except in the outer disk where  gas gravity dominates over stellar gravity.
If instead a radially varying scale height is considered, the full expression we recovered from the OK22 model, P$_{DE}^{1d,2d}$, as well as P$_{BR}^{1d,2d}$ , predicts a midplane pressure which differs by less than 20$\%$  than the true pressure. However, P$_{hy}^{1d}$ approximates P$_{true}$   better than P$_{DE}^{1d}$ and P$_{BR}^{1d}$. Notice in fact that their ratio with P$_{true}$  is not radially constant, opposite to the P$_{hy}^{1d}$ to P$_{true}$ ratio. Moreover P$_{BR}^{1d}$ uses an inconsistent and underestimated  stellar scale height (the one relative to a one component only layer) and this gives a lower pressure than a more self-consistent treatment would imply. The more correct treatment of the stellar scale height in P$_{BR}^{2d}$ underlines in fact how P$_{BR}$  predicts a  higher pressure than P$_{true}$, especially in the outer disk. We underline  that P$_{BR}^{2d}$ gives very similar values to P$_{DE}^{2d}$ only close to the center of M33 where gas is less relevant, but  they diverge further out. The dynamical equilibrium pressure for a radially varying scale height in a multicomponent layer , P$_{DE}^{2d}$,  reproduces better  P$_{true}$ than P$_{BR}^{2d}$. However, given the fact that we have slightly underestimated it, by using $h_s=h_g$  to compute the dark matter gravity in the stellar layer, we find that it does no better than P$_{hy}^{1d}$. These considerations apply whether we consider or not dark matter in the disk. Figure~\ref{p+pdm} in fact shows that the analytic expressions of the midplane pressure to the true pressure ratios are rather similar for the case with or without dark matter in the disk.

We conclude this Appendix by summarizing how to recover the dark matter density for a generic galaxy, for which the results of the dynamical analysis of the rotation are not know i.e. the concentration and halo mass of the dark matter components for the NFW halo model to infer the radial density distribution. Starting from the known stellar mass of the galaxy, we can recover the dark matter halo mass at virial radius from abundance matching \citep{2010ApJ...710..903M}. Using the halo concentration - halo mass relation \citep{2014MNRAS.441.3359D} one can infer the concentration parameter. From these relations, for a given stellar mass in units of M$_\odot$ one can recover the dark matter scale radius, r$_s$ in kpc and  the dark matter density at the scale radius $\rho_{dm}^{r_s}$ in M$_\odot$~pc$^{-3}$ as

\begin{equation}
{\hbox{log}}(r_s) \simeq 0.22\ {\hbox{log}}(M_s) -0.97  \qquad {\hbox{for\ }} M_s \le 5\ 10^{10}~M_\odot
,\end{equation}
\begin{equation}
{\hbox{log}}(\rho_{dm}^{rs})\simeq -0.12\ {\hbox{log}}(M_s) -1.70 \qquad {\hbox{for\ }} M_s \le 5\ 10^{10}~M_\odot
,\end{equation}
\begin{equation}
{\hbox{log}}(r_s) \simeq  0.98\ {\hbox{log}}(M_s) -9.09  \qquad {\hbox{for\ }} M_s > 5\ 10^{10}~M_\odot
,\end{equation}
\begin{equation}
{\hbox{log}}(\rho_{dm}^{rs})\simeq -0.52\ {\hbox{log}}(M_s) +2.56 \qquad {\hbox{for\ }} M_s > 5\ 10^{10}~M_\odot
.\end{equation}

 With these two parameters the dark matter density profile for a NFW dark halo a function of the distance from the galaxy center is

\begin{equation}
\rho_{dm}^{NFW}(r) = {4 \rho_{dm}^{r_s} \over {r\over r_s} (1+{r\over r_s})^2 }
.\end{equation}

\noindent
Given the stellar mass of M33 M$_s^{33}= 4.8\ 10^9$~M$_\odot$ \citep{2003MNRAS.342..199C} we can estimate from these relations for example  that the scale radius of M33 halo is   15.2~kpc and that the density at scale radius is 0.0012~M$_\odot$ pc$^{-3}$. With these predictions the dark matter density in the midplane at R=4~kpc  would be $\rho_{dm}$(R=4)=0.012~M$_\odot$ pc$^{-3}$. With the more accurate parameters from the dynamical analysis of the rotation curve  $\rho_{dm}^{RC33}$(R=4)=0.0094~M$_\odot$ pc$^{-3}$. This differs by less than 20$\%$ from the value estimated  from the above formulae. As an additional example let us consider the M31 halo. The dynamical analysis of the rotation curve predict a halo mass of 1.2 $10^{12}$~M$_\odot$ with a concentration C=12, which implies that at R=10~kpc $\rho_{dm}^{RC31}$(R=10)=0.0053~M$_\odot$ pc$^{-3}$ \citep{2010A&A...511A..89C}. Given the stellar mass of M31, M$_s^{31}=10^{11}$~M$_\odot$ the above relations predict for the star--forming ring at 10~kpc $\rho_{dm}(R=10)$=0.0094~M$_\odot$~pc$^{-3}$. There is about a factor 2 difference with $\rho_{dm}^{RC31}$ due to the fact that M31 does not lie along the stellar-to-halo mass relation, having a smaller halo mass than what abundance matching relations predicts, and a scale radius of 23~kpc, less than half of what the above relations predicts. For these reason we suggest to use the dark matter halo parameters from dynamical curve analysis when these are available \citep[e.g.][]{2020ApJS..247...31L}. This consideration applies especially to galaxies which are offset from the abundance match relation, which is the case for M31 \citep{2019A&A...626A..56P}.

\section{Radial variations and the Toomre Q parameter }
\label{apprad}

In this Appendix, we show radial variations of several physical quantities across the star-forming disk of M33  with the aim to see if we can distinguish different regimes; namely, if there is some transition radius beyond which radial trends of physical quantities and star formation regulation change. In analyzing  CO and HI spectral data for this galaxy \citet{2019A&A...622A.171C} have pointed out some dynamical differences between the inner and outer star-forming regions of this galaxy, namely: $(i)$ the corotation radius of the spiral pattern is located at  4.7$\pm$0.3~kpc, and the sonic point, defined as the region where the velocity difference between the spiral pattern speed and the rotational velocity equals the sound speed, is at 3.9~kpc.  $(ii)$ the mean value of GMC mass increases as the radius decreases for R$<$4~kpc, while it stays constant beyond this radius; the CO emission fraction not related to GMCs increases dramatically beyond 4~kpc indicating a pervasive population of low-mass bound clouds or the dominance of diffuse molecular gas. $(iii)$ the number density of luminous mid-IR sources (with F$_{24\mu m} >$5~mJy) decreases sharply beyond 4~kpc. Clearly inside corotation, and especially inside the sonic point,  the spiral pattern is able to collect gas condensations into larger clouds and trigger star formation through shocks. At larger radii the two main arms dissolve while thin elongated filaments, resulting naturally from disk instabilities \citep{2018MNRAS.478.3793D}, survive and provide a fertile environment for star formation. 

 \begin{figure} 
\includegraphics [width=9.0 cm]{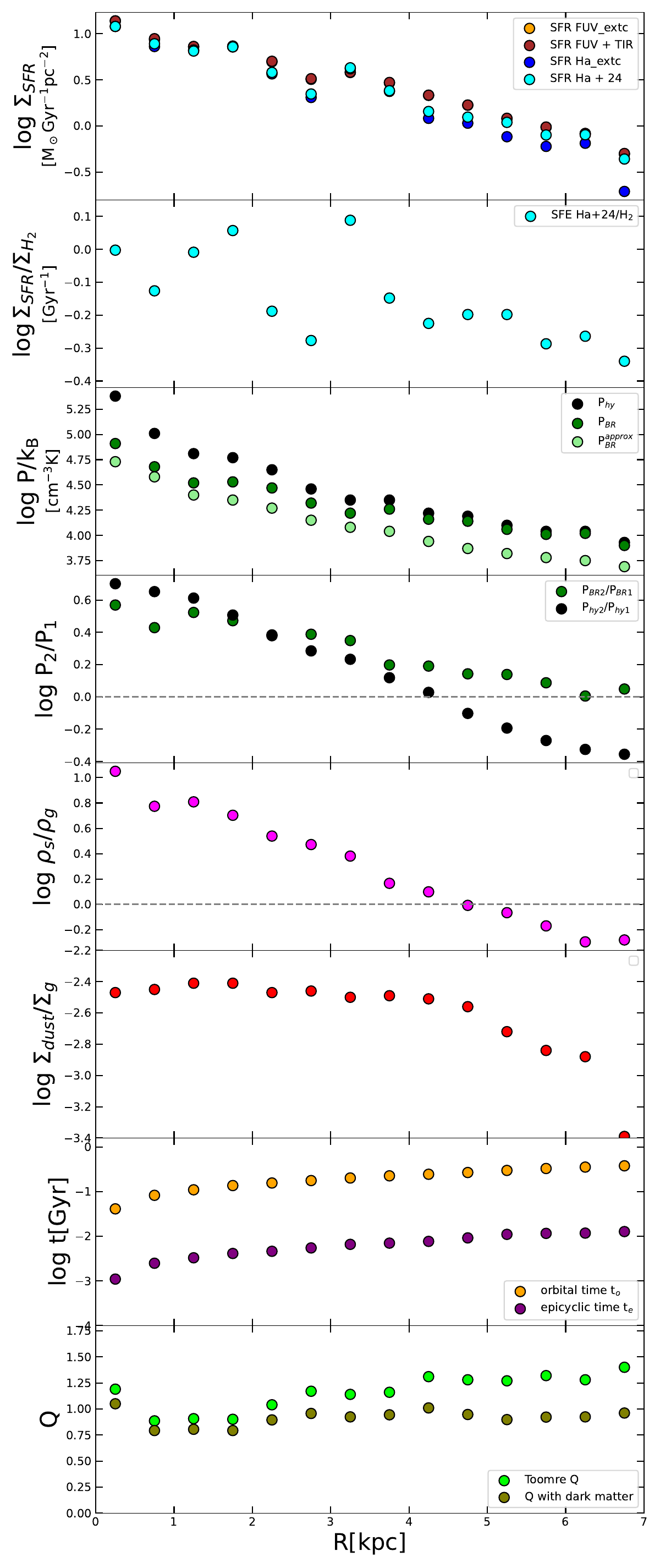}
\caption{ Radial averages in M33 as a function of galactocentric distance R.    From top to bottom:\ Star formation rate surface density, $\Sigma_{SFR}$, using various indicators (see Appendix~\ref{appfuv} for details), the effective SFE, $\Sigma_{SFR}/\Sigma_{H_2}$, the midplane pressures P$_{hy}$, P$^0_{BR}\equiv$ P$^*_{DE}$, and P$^{*,approx}_{DE}$, the ratio between the stellar and gaseous pressure terms, P$_2$/P$_1$, in P$^*_{DE}$ and P$_{hy}$, the  stellar to gas volume density ratio $\rho_{s}/\rho_{g}$, the dust  to total gas mass surface density ratio $\Sigma_{dust}/\Sigma_g$ , the orbital, and epicyclic times t$_o$, t$_e$,  and    the Toomre parameter Q for a thin disk with or without considering the dark matter gravity.  The gray dashed lines indicate unity ratios.}
\label{sfr_time}
\end{figure}

In Figure~\ref{sfr_time}  we show additional radial profiles to those already shown in  Figure~\ref{scale_r}. The average radial decline of $\Sigma_{SFR}$ is similar if we consider the combination of H$\alpha$ and 24$\mu$m emission or that of FUV and TIR emission, or also if we correct H$\alpha$ or FUV emission for extinction.
The dispersion around the mean values however depends on the SFR tracer and this might affect  local relations  (see Appendix~\ref{appfuv}). 

The effective SFE, defined as the  ratio $\Sigma_{SFR}/\Sigma_{H_2}$,  is the inverse of the molecular gas consumption time for a sufficiently large region.  The mean efficiency oscillates between 1.2 and 0.8~Gyr$^{-1}$ inside 4~kpc  and it steadily declines radially beyond 4~kpc. The  hydrostatic pressure  decreases radially by about 1.5 orders of magnitudes, similarly to $\Sigma_{SFR}$, going from the center to the edge of the star-forming disk. The simplified dynamical equilibrium pressure P$_{DE}^*$, computed for a radially constant scale height, varies instead only by one order of magnitude. The radial decline would be faster by considering more accurate P$_{DE}$ expressions with a radially varying gas scale height. We also show the radial decline of the approximation used for P$_{DE}^*$,  P$_{DE}^{*,approx}$, by setting $\Sigma_{HI}=7$~M$_\odot$~pc$^{-2}$ and c$_g^0=11$~km~s$^{-1}$ uniformly throughout the disk.
Values of P$_{DE}^{*,approx}$ (see Section~\ref{secphy} or Appendix~\ref{appsh}), adopted independently of galaxy type when atomic data at adequate resolution is lacking and representative of  the range  found in disc galaxies for the cold gas \citep{2021MNRAS.503.3643B,2013AJ....146..150C,2008AJ....136.2782L} trace well P$_{DE}^*$ in the inner disk. In Figure~\ref{sfr_time} we see however that for an atomic gas dominated galaxy, like M33, it underestimates  the true pressure beyond 3.5~kpc. 

The radial decrease in the hydrostatic pressure is less uniform across the disk than that of $\Sigma_{SFR}$,  being steeper  in the inner region and flatter at larger radii. This is due to the presence of two terms
in the hydrostatic pressure:  P$_{hy2}$, given by a combination of $\Sigma_s$   and $\Sigma_g$, which decreases consistently as R increases because of the fast $\Sigma_s$ radial decline,  and P$_{hy1}$
which has a slower radial decline, given the shallow radial profile of the gas surface density with the dark matter contribution which becomes more relevant at large radii. Comparing  the two terms  we see that  beyond 4.2~kpc P$_{hy1}$ dominates over P$_{hy2}$ changing the slope of the hydrostatic pressure radial decline. 

The stellar volume density,  $\rho_{s}$, is  much higher than the gas volume density in the inner disk while it becomes lower than $\rho_{\rm g}$  for R$> 4 $~kpc. This limiting radius is also the boundary of the region where  the weight of the  stellar disk   dominates over the gas and dark matter vertical weight. 
The mean dust-to-gas mass surface density ratio is nearly constant with radius  for R$<$4~kpc and declines as the galactocentric radius increases beyond 4~kpc.  Some caution should be used  in interpreting the radial trend of $\Sigma_{dust}/\Sigma_{gas}$  shown in Figure~\ref{aper_r}. The dust mass surface density has a shallow radial decline out to about 5 kpc. Beyond this radius the emission in the far-IR band decreases such that FIR map sensitivities are not enough to fully cover the whole disk extent. We have set equal to zero the dust mass surface density of pixels with a low signal to noise in one of the FIR bands. For these reason, by including all the pixels during radial averages we recover a lower limit for the mean $\Sigma_{dust}$ and $\Sigma_{dust}$-to-$\Sigma_{gas}$ ratio at R$>$5~kpc.   However the radial decline  is confirmed when we neglect areas with low dust surface density  and whose $\Sigma_{dust}$ has been set to zero, as we will show later in this Appendix.  

In Figure~\ref{sfr_time} we show also the mean radial variations of the epicyclic and orbital time, t$_e$ and t$_o$, respectively,  defined as follows

\begin{equation}
t_e(R)={1\over \kappa_e(R)}=\Bigl[{R{d\Omega^2(R)\over dR}+4\Omega^2(R)}\Bigr]^{-0.5} \qquad t_o(R)={2\pi\over {\Omega(R)} }
.\end{equation}

\begin{figure} 
\includegraphics [width=9.0 cm]{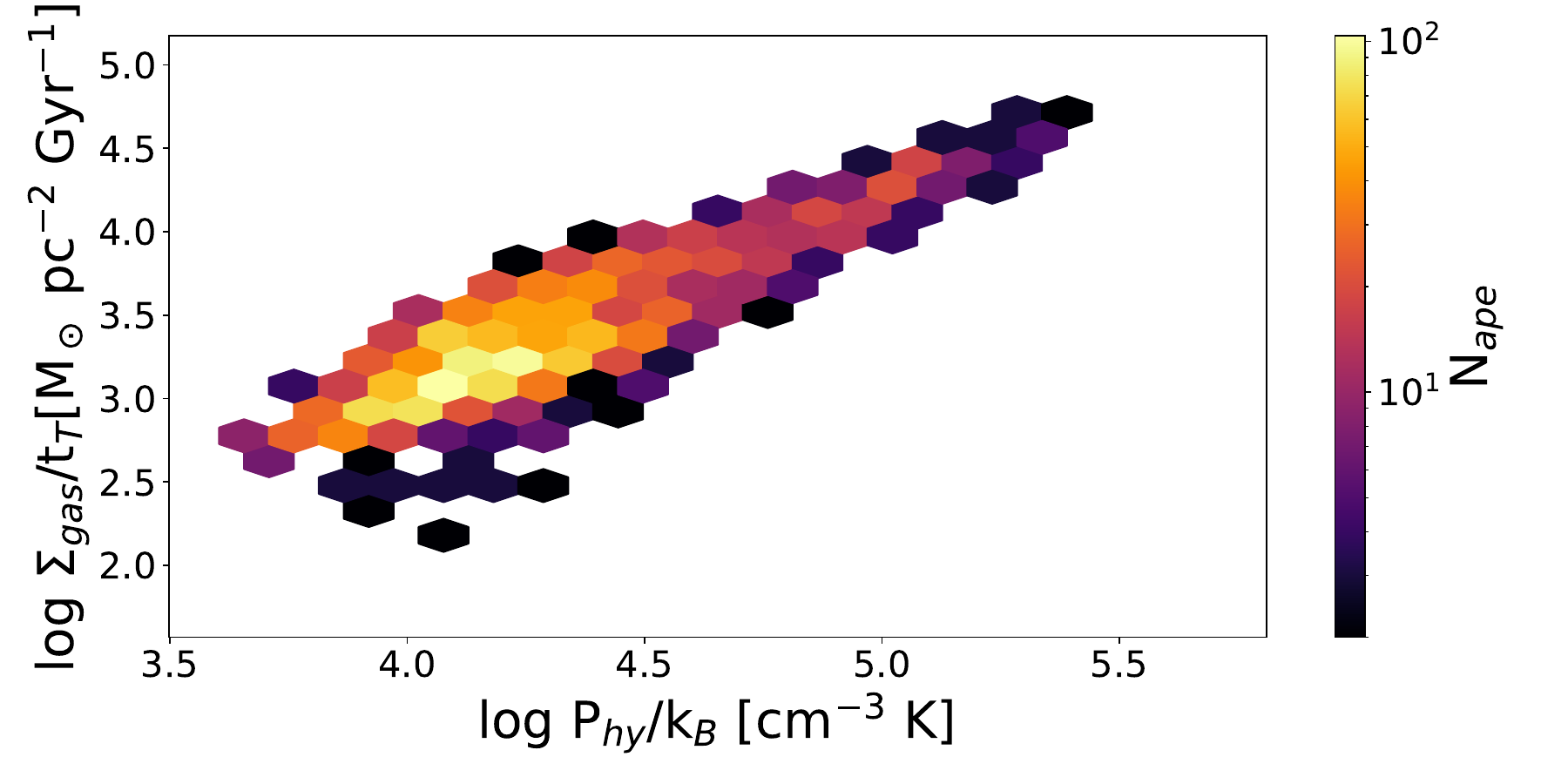}
\caption{Strong correlation between the hydrostatic pressure, P$_{hy}$, and the condensation parameter $\Sigma_{gas}$/t$_T$ for regions where Q$<1$.} 
\label{toomre}
\end{figure}

\begin{figure} 
\includegraphics [width=9.0 cm]{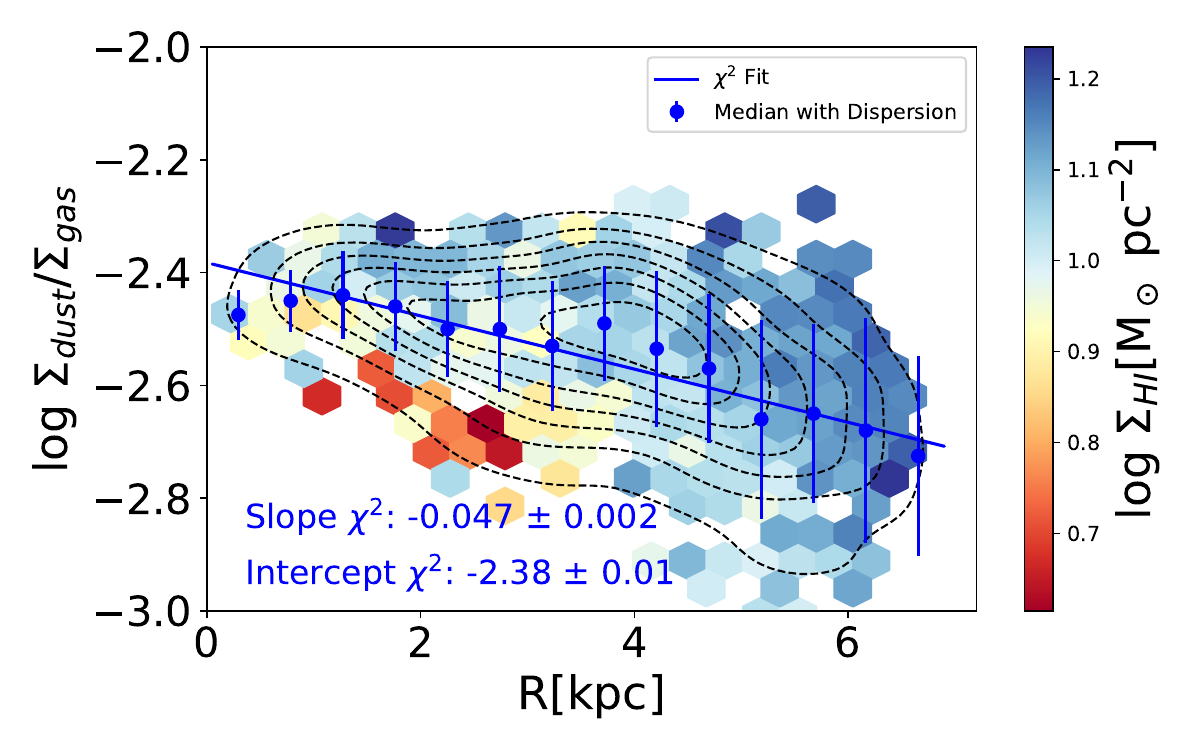}
\caption{Dust-to-gas mass surface density ratio $\Sigma_{dust}/\Sigma_{gas}$  in  apertures with R$_a$=30~arcsec as a function of galactocentric radius R. Hexagons are color coded according to the atomic gas surface density $\Sigma_{HI}$,  given by the colorbars,  The blue line shows the linear fit to the data as given by  $\chi^2$ minimization.  After radial binning, the median values and their dispersion are shown with filled blue circles and vertical blue bars, respectively. The dashed lines indicate the density of data points.}
\label{aper_r}
\end{figure}

\noindent
In the above equations $\kappa_e$  is the epicyclic frequency and  $\Omega$ the angular speed at a given galactocentric radius R. These times  have  similar radial trends and provide an estimate of the time it takes for perturbations related to a rotating disk to grow. The epicyclic time is of the order of a few million years in the inner disk, increasing radially to reach 10 million years at the edge of the star-forming disk. We have also computed the Toomre parameter Q for a two component disk, having stars and gas as in \citet{2011MNRAS.416.1191R,2011ApJ...737...10E,2016MNRAS.456.2052I}.  The Q values for a thin disk range between 0.9 and 1.4 for R$<7$~kpc increasing to larger values beyond this radius. Considering the dark matter contribution to the gas gravity  we have a more uniform Q$\simeq 1$, although it is unclear if dark matter helps the formation of disk instabilities given its large dispersion (in the range 30-70~km~s$^{-1}$ for M33 halo). To account for the disk finite thickness we should furthermore multiply the Q values by a factor 1.44  \citep{1992MNRAS.256..307R}  which would imply $1<$Q$<2$ and hence a marginally stable disk.  However, given the uncertainties on the radial velocity dispersions and their corrections for the specific heats \citep{1996ASSL..209..467E},  together with the possibility that marginally stable disks might still  form elongated thick filaments due to azimuthal perturbations \citep{2018MNRAS.478.3793D}, we estimate  the condensation Toomre timescale,  t$_T$, as in  \citet{2015MNRAS.448.1007B}: 

\begin{equation}
t_T= \kappa_e^{-1}\Biggl({1\over Q^2} -1 \Biggr)^{-1/2}
,\end{equation}

\noindent
 considering only regions for which the conditions for a thin unstable disk with dark matter are satisfied.   The Toomre condensation parameter, which is the ratio between the gas mass surface density and the characteristic Toomre timescale, $\Sigma_{gas}$/t$_T$, can also be used to examine star formation drivers possibly  related to the large scale disk motion. This parameter  is correlated with P$_{hy}$, as shown in Figure~\ref{toomre}. In this work we limit the analysis of the Toomre condensation parameter to RF regression  (see Section 4.4) for regions with Q<1, but results suggest further analysis along these lines taking into account the non negligible disk thickness.
 
Using aperture radius R$_a$=30~arcsec we show in Figure~\ref{aper_r}  the dust-to-gas mass surface density ratio $\Sigma_{dust}/\Sigma_{gas}$  in the apertures as a function of galactocentric radius. The blue line shows the linear fit to the data as given by the least-squares fit (results in this case are coincident with orthogonal regression fit).  Hexagons are color coded according to atomic gas surface density related to the observed dispersion for each radius. After radial binning, the median values and their dispersion  are shown with filled blue circles and vertical blue bars. We notice that the dust to gas ratio is higher in gas rich regions, with  atomic gas filaments tracing  dusty regions from the inner to the outer disk of M33. The shallow slope of the dust-to-gas ratio radial profile is similar to the gas metallicity gradient slope in the inner disk \citep[e.g.][]{2010A&A...512A..63M} (including the drop in the central region). In the outer disk the dust-to-gas ratio has a high dispersion and lower values on average.   The radial fits of the dust-to-gas ratio and of other variables such as  P$_{hy}$ improve using two linear functions, one for the inner region  and the other for the outer region. However by leaving the transition radius between inner and outer region as a free parameters in  fitting the data,  we are not  able to recover solid results. Often the resulting transition radius is closer to the edge of the star-forming disk (which is where we have numerous data points) or it is driven by outliers. The results and considerations expressed in this Appendix support the definition of the inner star-forming disk for R$\le 4$~kpc, and the outer star forming disk for $4<$R$\le 7$~kpc.

\section{Variations of star formation tracer and  other parameters}
\label{appfuv}

In Section~3, we have introduced the combined H$\alpha$ line and 24$\mu$m emission star formation indicator and used this throughout the main paper because we are interested in the recent star formation. Here we  use H$\alpha$ and FUV maps to trace the SFR across the M33 disk after applying extinction corrections, or we  combine the  FUV emission  with the  total infrared flux. The visual extinction A$_v$ has been estimated using  the far-infrared emission as described by \citet{2001PASP..113.1449C} through the following equation:
\begin{equation}
A_v = C \ 1.76\ {\hbox{log}} \Bigl({L_{TIR} \over 1.68 L_{FUV}} + 1 \Bigr)
,\end{equation}
where the total infrared (hereafter TIR) luminosity  has been computed following Eq.(1) of \citet{2009A&A...493..453V}. The value of C depends on whether we consider extinction for the stellar continuum or  for the ionized gas: C=0.44 for stellar continuum while for the ionized gas, more clustered around star-forming regions C=1 \citet{2001PASP..113.1449C}. For the ionized gas in M33 we use C=0.7 instead of C=1, as discussed in \citet{2009A&A...493..453V},  because of the large diffuse gas fraction in the disk.  A standard extinction curve has been considered with A$_v$=3.1 E(B-V), and following \citet{2005ApJ...619L..55S} we computed A$_{FUV}$=8.29 E(B-V) . The extinction corrected H$\alpha$ and FUV luminosity has been converted into a SFR as in \citet{2012ARA&A..50..531K}:

\begin{equation}
SFR [M_\odot yr^{-1} ] = 4.47\times 10^{-44}  L_{FUV}^{cor} [erg s^{-1}] 
,\end{equation}

\begin{equation}
SFR [M_\odot yr^{-1} ] = 5.37\times 10^{-42}  L_{H\alpha}^{cor} [erg s^{-1}] 
.\end{equation}

\noindent
 The use of extinction-corrected H$\alpha$ emission as SFR tracer gives a lower SFR density for the outer  regions. In this case the relation with P$_{hy}$ has a larger scatter and the relation steepens  at low pressures.  A faster decline of extinction corrected H$\alpha$ luminosity with respect to FUV  luminosities has been reported for M33 regions also  by \citet{2010A&A...510A..64V}, where the authors discuss the IMF incompleteness as possible driver of this effect. We suggest here that  the use of the estimated TIR emission to correct H$\alpha$ for extinction  might contribute to this, because of the pervasive  diffuse ISM component. The absorption of ionizing photons  by dust in the HII regions, and the hot dust component make the extinction corrections of H$\alpha$ line through the TIR emission not ideal.  Moreover, as shown by \citet{2015A&A...578A...8B}, differential reddening depends on the physical scale and on the specific SFR. 
 
 We also combine  FUV emission with the TIR emission to estimate  $\Sigma_{SFR}$ following \citet{2011ApJ...741..124H,2012ARA&A..50..531K}:

\begin{equation}
SFR [M_\odot\ yr^{-1} ] = 4.47\times 10^{-44} \Bigl( L_{FUV} + 0.46 L_{TIR} \Bigr)
,\end{equation}

\noindent
where  luminosities are in erg~s$^{-1}$. The values of $\Sigma_{SFR}$ are very similar to those obtained for the extinction corrected FUV or a combination of FUV and 24$\mu$m emission.
Young stars over 100-300 Myr are more spread throughout the ISM and their extinction is closely related to the diffuse dust.. Tracing the SFR with FUV emission gives a higher total SFR for this galaxy \citep{2009A&A...493..453V} and for most of the regions. On average the ratio of $\Sigma_{SFR}$  traced by H$\alpha$+24$\mu$m   to that traced by any FUV  tracer is $<1$ (with an average value of -0.15 dex). When  L$_{H\alpha} <3\times 10^{37}$~erg~s$^{-1}$, which corresponds to log $\Sigma_{SFR}\simeq 0.5$~M$_\odot$~pc$^{-2}$~Gyr$^{-1}$ for R$_{a}$=30~arcsec, the ratio of H$\alpha$-to-FUV luminosities is expected to decreases due to IMF incompleteness, as fainter regions are sampled \citep{2009A&A...495..479C,2010A&A...510A..64V}. However, the observed decreasing trend  does not occur only for luminosities below a certain threshold, and it has a large dispersion (larger than that reported by \citet{2010A&A...510A..64V} using extinction corrected H$\alpha$). It seems unlikely that is only the IMF incompleteness the driver of this trend and other processes might play a role, such as photon leakage from HII regions . Moreover, a similar decreasing trend is observed when using  R$_{a}$=120~arcsec, for which only a few regions have L$_{H\alpha} <3\times 10^{37}$~erg~s$^{-1}$ and are affected by IMF stochasticity. A possibility is that the enhanced FUV emission is due to a strong star formation episode in M33 which happened 300-500~Myr ago as reported by deep photometric studies of the extended outer disk \citep{2011A&A...533A..91G}.

\begin{figure} 
\includegraphics [width=9. cm]{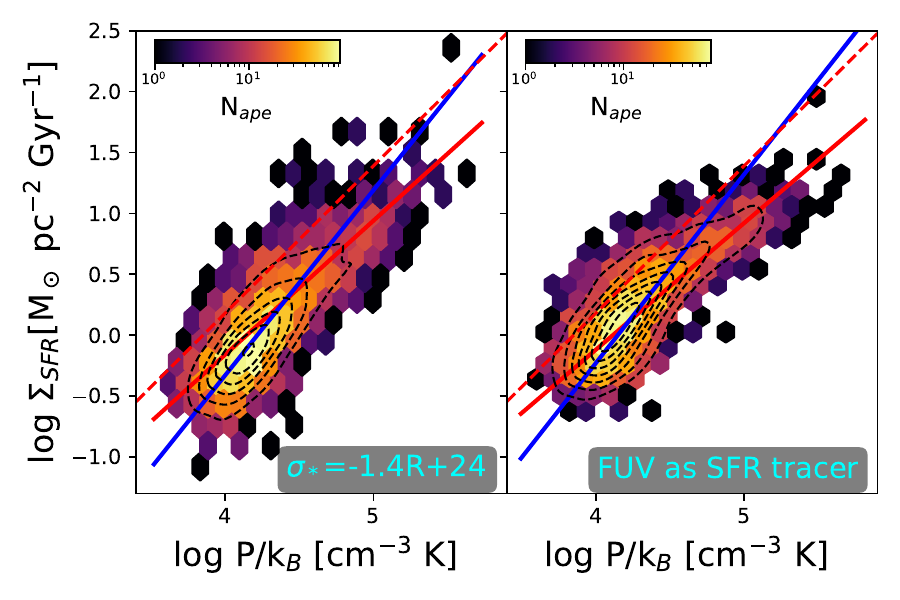}
\caption{Relation between P$_{hy}$ and $\Sigma_{SFR}$  is shown by varying some of our working assumptions. In the left panel we consider a possible negative radial gradient in the stellar velocity dispersion, and in the rightmost panel we trace star formation using FUV luminosities.  Color coding and contours indicate the plane.}
\label{aperD1}
\end{figure}

Our aim here is to check that our main conclusions hold computing $\Sigma_{SFR}$ from FUV tracer. To this purpose we re-run some of our scaling relation and RF analysis. The scaling relations for $\Sigma_{SFR}$ are all flatter than those shown in Section~5 for H$\alpha$+24$\mu$m tracer. This is because the star formation rate surface density from the FUV tracer has a shallower  decline with the luminosity of the region  than the H$\alpha$+24$\mu$m SFR tracer. However, also for this tracer the $\Sigma_{SFR}$--P$_{hy}$ relation is tight and steepens in the outer disk, as shown in Figure~\ref{aperD1}. 

For the outer disk the use of the extinction corrected FUV  indicator as star formation tracer, less affected by stochastic variations in the IMF, doesn't give a better agreement between the predicted and the observed $\Sigma_{SFR}$. With the FUV as SFR tracer, we noticed a somewhat higher relevance is found for the dust surface density. Given the generally low value of R$^2$, and the dependence of the relative importance inferred from RF regression from SFR tracer, star formation driver in the outer disk of M33  would benefit from further investigation.

We also considered a possible radial gradient in the stellar velocity dispersion, as that suggested by measurements relative to the old stellar population which can be described by c$_s$[km~s$^{-1}$] = -1.4 R[kpc] + 24  \citep{2022AJ....163..166Q}. With this negative radial gradient, we do have a few regions in the outer disk where the assumption for the stellar scale height being higher than the gas scale height is not verified (because the stellar layer is less extended than in the case of a radially constant dispersion). The main conclusions about the  scaling relations discussed in this paper however still hold when using this velocity dispersion gradient. Slopes get slightly steeper when P$_{hy}$ is involved, but they remains within the uncertainties shown in Table~1.

To confirm the  results of RF regression we have also carried out the RF analysis by changing the condition for the inclusion of apertures with missing pixels in the dust surface density mass map. We have varied the condition for the inclusion from the standard case, at least 50$\%$ of the pixels with determined dust mass, to the limiting case, 100$\%$ of the pixels with determined dust mass. For all cases the dust mass surface density has been computed by considering only the pixels in the aperture for which the dust mass has been determined.  We did not find any reasonable variations of our results.

 \end{appendix}

\end{document}